 \documentclass[twocol]{ametsoc}

\journal{jas}

\bibpunct{(}{)}{;}{a}{}{,}

\usepackage{microtype}
\usepackage{doi}

\usepackage{hyperref}
\hypersetup{
	breaklinks,
	colorlinks=true,
	linkcolor=blue,
	citecolor=purple,
	bookmarks=true,
	bookmarksopen=true,
	bookmarksopenlevel=2,
	pdfstartpage={1},
	pdfstartview={FitH},
	pdfview={FitH 0},
	pdfauthor={N A Bakas, N C Constantinou and P J Ioannou},
	pdftitle={S3T stability of the homogeneous  state of barotropic beta-plane turbulence},
 }

\def\deg{$^{\circ}$N}

\def\Uv{\boldsymbol{U}}

\newcounter{saveeqn}%

\newcommand{\be}{\begin{equation}}
\newcommand{\ee}{\end{equation}}
\newcommand{\bdm}{\begin{equation*}}
\newcommand{\edm}{\end{equation*}}
\newcommand{\bea}{\begin{eqnarray}}
\newcommand{\eea}{\end{eqnarray}}

\newcommand{\iintinf}{\iint\limits_{-\infty}^{~~~+\infty}}
\newcommand{\partialf}[2]
{
 \ifthenelse{\equal{#1}{}}{\frac{\partial}{\partial #2}}{\frac{\partial #1}{\partial #2}}
}

\newcommand{\sgn}{\mathop{\mathrm{sgn}}}
\newcommand{\real}{\mathop{\mathrm{Re}}}
\newcommand{\imag}{\mathop{\mathrm{Im}}}

\renewcommand{\(}{\left(}
\renewcommand{\)}{\right)}
\renewcommand{\[}{\left[}
\renewcommand{\]}{\right]}
\newcommand{\<}{\left\langle}
\renewcommand{\>}{\right\rangle}

\newcommand{\e}{\varepsilon}
\newcommand{\Del}{\Delta}

\renewcommand{\d}{\delta}

\newcommand{\df}{\textrm{d}}

\newcommand{\la}{\lambda}
\newcommand{\s}{\sigma}
\renewcommand{\b}{\beta}

\newcommand{\z}{\zeta}

\renewcommand{\i}{\mathrm{i}}

\newsavebox{\astrutbox}
\sbox{\astrutbox}{\rule[-5pt]{0pt}{20pt}}

\def\bit{\vphantom{\dot{W}}}

\def\st{\sin{\vartheta}}

\def\x{\chi}
\def\om{\omega}

\def\thet{\vartheta}

\def\dvz{\d\,\overline{v'\z'}}

\def\sR{s_r}
\def\sR{f_r}

\def\Gcal{\mathcal{G}}
\def\Dcal{\mathcal{D}}
\def\Db{\Dcal_2}
\def\Dbj{\Dcal_{2,j}}
\def\Ocal{\mathcal{O}}
\def\Fcal{\mathcal{F}}
\def\Acal{\mathcal{A}}
\def\Tcal{\mathcal{T}}
\def\Lcal{\mathcal{L}}
\def\Rcal{\mathcal{R}}

\def\FcalR{\mathcal{F}^{(R)}}
\def\FcalR{\mathcal{F}}
\def\Ncal{\mathcal{N}}
\def\deg{^\circ}

\def\xv{\mathbf{x}}
\def\bv{\bm{\b}}
\def\Uv{\mathbf{U}}
\def\uv{\mathbf{u}}

\def\nv{\mathbf{n}}
\def\kv{\mathbf{k}}

\def\dU{\delta\tilde{U}}

\def\dZ{\delta\tilde{Z}}
\def\dC{\delta\tilde{C}}

\def\pv{\mathbf{p}}

\def\spa{\vphantom{\frac1{(2\pi)^2}\frac{\Ncal\,\Dcal_0}{ \Dcal_0^2 + \b^2\,\Db^2 }}}
\def\spb{\vphantom{\int\limits_{\thet_{j-1}+\d}^{\thet_j-\d\thet} }}

\newcommand{\tav}[1]{\Tcal\[\,#1\,\]}
\def\dvz{\d\< v' \z'\>}

\def\f{f}
\def\fr{\sR}
\def\zhat{\hat{\mathbf{z}}}
\def\nablav{\bm\nabla}

\title{S3T stability of the homogeneous state of barotropic beta-plane turbulence}

    \authors{Nikolaos A. Bakas, Navid C. Constantinou\correspondingauthor{Navid Constantinou, University of Athens, Department of Physics, Section of Astrophysics, Astronomy and Mechanics, Build IV, Office 32, Panepistimiopolis, 15784 Zografos, Athens, Greece.} and Petros J. Ioannou}

     \affiliation{Department of Physics, National and Kapodistrian University of Athens, Athens, Greece}

\email{navidcon@phys.uoa.gr}

\abstract{Zonal jets and non-zonal large-scale flows are often present
in forced-dissipative barotropic turbulence on a beta-plane. The
dynamics underlying the formation of both zonal and non-zonal coherent
structures is investigated in this work within the statistical framework
of Stochastic Structural Stability Theory (S3T). Previous S3T studies
have shown that the homogeneous turbulent state undergoes a bifurcation
at a critical parameter and becomes inhomogeneous with the emergence of
zonal and/or large-scale non-zonal flows and that these statistical
predictions of S3T are reflected in direct numerical simulations. In
this paper, we study the dynamics underlying the S3T statistical
instability of the homogeneous state as a function of parameters. It is
shown that for weak planetary vorticity gradient, $\b$, both zonal jets
and non-zonal large-scale structures form from upgradient momentum
fluxes due to shearing of the eddies by the emerging infinitesimal
flow. For large $\b$, the dynamics of the S3T instability differs for
zonal and non-zonal flows but in both the destabilizing vorticity
fluxes decrease with increasing $\b$. Shearing of the eddies by the mean
flow continues to be the mechanism for the emergence of zonal jets while
non-zonal large-scale flows emerge from resonant and near resonant triad
interactions between the large-scale flow and the stochastically forced
eddies. The relation between the formation of large-scale structure
through modulational instability and the S3T instability of the
homogeneous state is also  investigated and it is shown that the modulational
instability results are subsumed by the S3T results.}

\begin{document}

\maketitle


\section{Introduction}

Atmospheric turbulence is commonly observed to be organized into slowly
varying large-scale structures such as zonal jets and coherent vortices.
Prominent examples are the banded jets and the Great Red Spot in the
Jovian atmosphere \citep{Ingersoll-90,Vasavada-and-Showman-05}.
Laboratory experiments as well as direct numerical simulations of
turbulent flows have shown that these coherent structures appear and
persist for a very long time despite the presence of eddy mixing
\citep{Vallis-Maltrud-93,Weeks-etal-1997,Read-etal-2004,Espa-etal-2010,
DiNitto-etal-2013}.

A model that exhibits many aspects of turbulent self-organization into
coherent structures yet is simple enough to extensively investigate, is
a barotropic flow on the surface of a rotating planet or on a beta-plane
with turbulence sustained by random stirring. Numerical simulations of
this model have shown that robust zonal jets coexist with large-scale
westward propagating coherent waves
\citep{Sukoriansky-etal-2008,Galperin-etal-2010}. These waves were found
to either obey a Rossby wave dispersion, or form non-dispersive packets
that are referred to as satellite modes \citep{Danilov-04} or
zonons~\citep{Sukoriansky-etal-2008}. In addition, the formation of
these coherent structures was shown to be a bifurcation phenomenon. As
the energy input of the stochastic forcing is increased, the flow
bifurcates from a turbulent, spatially homogeneous state to a state in
which zonal jets and/or non-zonal coherent structures emerge and are
maintained by turbulence
\citep{Bakas-Ioannou-2013-prl,Constantinou-etal-2014}. In this work, we
will address the eddy--mean flow dynamics underlying the emergence of
both zonal and non-zonal structures.

Since organization of turbulence into coherent structures involves
complex nonlinear interactions among a large number of degrees of freedom, which
erratically contribute to the maintenance of the large-scale structure, an attractive
approach is to study the statistical state dynamics (SSD) of the turbulent flow, rather
than single realizations of the turbulent field. Recently, the SSD of barotropic and
baroclinic atmospheres has been studied by truncating the infinite hierarchy of cumulant
equations to second order. Stochastic Structural Stability Theory (S3T) is such a second
order Gaussian approximation of the full SSD, in which the third cumulant is parameterized as
the sum of a known correlation function and a dissipation term \citep{Farrell-Ioannou-2003-structural}.
This is equivalent to a  parametrization of the eddy--eddy nonlinearity as random forcing with the
required  dissipation to remove the energy injected by the forcing. Such a representation is strongly
supported by the results of previous studies. Linear inverse modeling studies showed that this
parametrization is the best linear representation of the eddy--eddy nonlinear interactions in planetary
turbulence \citep{DelSole-Farrell-1996,DelSole-96,DelSole-Hou-1999,DelSole-04}, while earlier
studies have shown that the turbulent transport properties (heat and momentum fluxes)
of the midlatitude transient climatology are accurately obtained as the stochastic response of the
large-scale flow to stochastic forcing \citep{Farrell-Ioannou-1994a,Farrell-Ioannou-1995,
Whitaker-Sardeshmukh-98,Zhang-Held-99}. In addition, \cite{Bouchet-etal-2013} have shown that in the
limit of weak forcing and dissipation, the formal asymptotic expansion of the probability density function
of the Euler equations around a mean flow that is assumed to only have a singular spectrum of modes, comprises of
the second order S3T closure with an additional stochastic term forcing the mean flow. Therefore, S3T formally
describes the statistical equilibrium mean flow and the eddy statistics in this case, as the additional stochastic
term only produces fluctuations around this statistical equilibrium.
Similar to the S3T closure of the full SSD is the CE2 closure in which the third order cumulant is neglected
without parameterization \citep{Marston-etal-2008, Marston-2010,Marston-2012,Tobias-Marston-2013}. It has
been shown that the predictions of S3T (or CE2) simulations  are reflected in corresponding  nonlinear
simulations \citep{OGorman-Schneider-2007,Srinivasan-Young-2012,Tobias-Marston-2013,Constantinou-etal-2014}.

The second order closure results in a
nonlinear, autonomous dynamical system that governs the evolution of
the mean flow and its consistent second order perturbation statistics.
Its fixed points define statistical equilibria, whose instability brings
about structural reconfiguration of the mean flow and of the turbulent
statistics. Previous studies employing S3T addressed the bifurcation from a
homogeneous turbulent regime to a jet forming regime in barotropic
beta-plane turbulence and identified the emerging jet structures as linearly unstable
modes to the homogeneous turbulent state equilibrium \citep{Farrell-Ioannou-2003-structural,Farrell-Ioannou-2007-structure,
Bakas-Ioannou-2011, Srinivasan-Young-2012,
Parker-Krommes-2013,Parker-Krommes-2014-generation}. The stability analysis of the
homogeneous equilibrium was further advanced with the introduction of the continuum
formulation in S3T theory by \cite{Srinivasan-Young-2012} who derived
a compact analytic expression for the growth rate and frequency of the unstable structures.
Interestingly, \cite{Carnevale-Martin-1982} using field theoretic techniques have arrived
at the same stability equation for the statistical description of fluctuations about a
homogeneous state.

Comparisons of the jet structure predicted by S3T theory  with direct numerical simulations have
shown that the structure of zonal flows that emerge in the nonlinear
simulations can be predicted by S3T \citep{Srinivasan-Young-2012,Tobias-Marston-2013,Constantinou-etal-2014}.
However, \cite{Srinivasan-Young-2012}  found quantitative differences
between the predictions of S3T regarding the bifurcation diagram for the
emergence of jets and the corresponding diagram obtained from the nonlinear
simulations, calling into question the validity of the S3T (or CE2) approximations when the mean flow
is very weak. \cite{Constantinou-etal-2014} demonstrated that this discrepancy was
due to the prior emergence of non-zonal coherent structures that modified the  background equilibrium
spectrum and showed that S3T predictions were accurate when this modification in the background
spectrum was accounted for.

\begin{figure}
\centerline{\includegraphics[width=19pc]{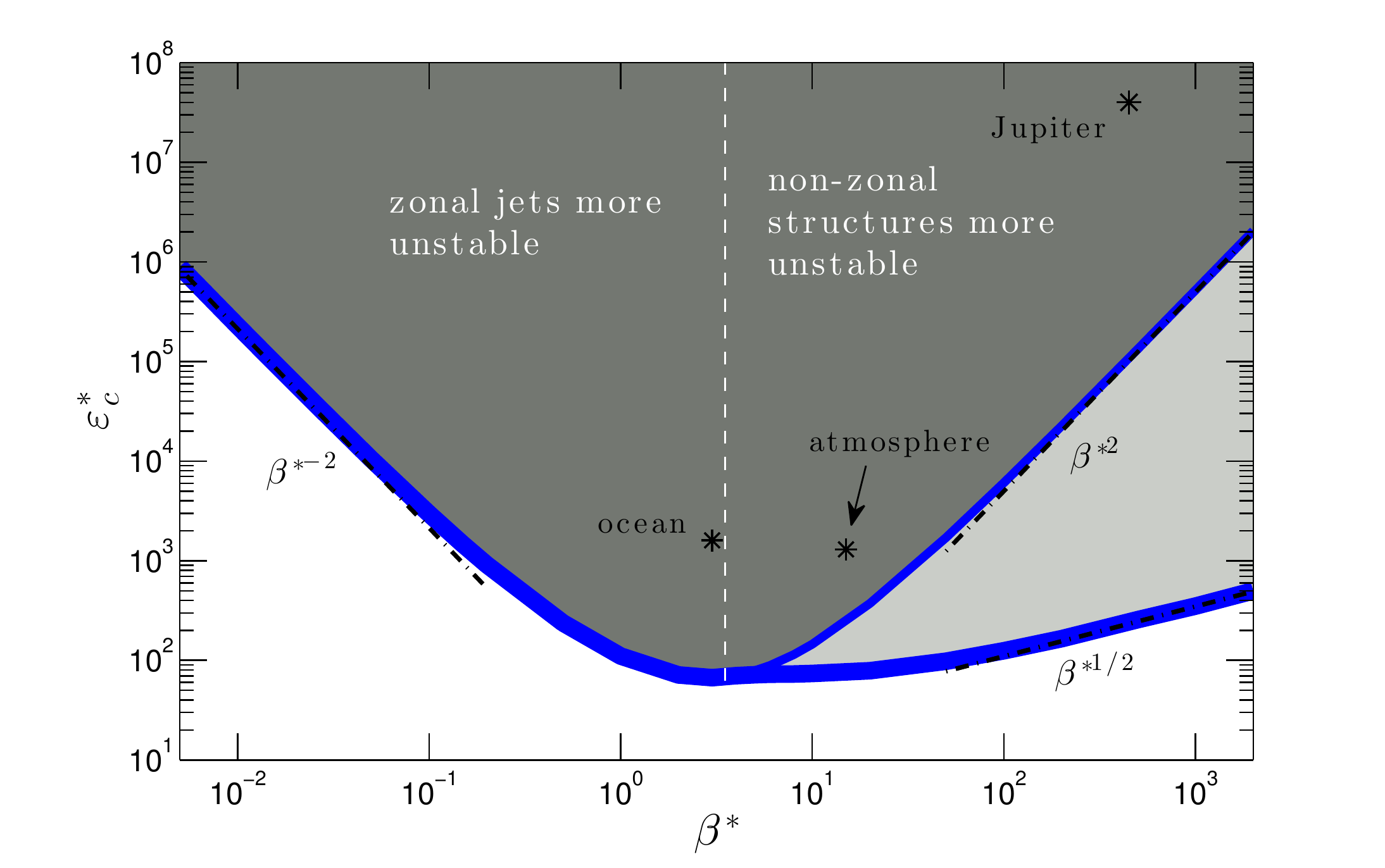}}
\caption{The critical energy input rate $\e_c^*$ for  the emergence of either zonal or non-zonal  large-scale structure (thick solid line)
and the critical $\e^*$  for the emergence of   zonal jets (solid line) as a function of the non-dimensional planetary vorticity gradient, $\b^*$,
when the stochastic forcing is isotropic. Jets or non-zonal structures emerge with the least energy input for $\b_{\min}^*\approx 3.5$.  For $\b^* \gg 1$ the critical input rate
for the emergence of jets increases as $\e_c^* \sim \beta^{* 2}$ and as $\e_c^* \sim {\b^*}^{1/2}$ for the emergence of non-zonal structures. In the light shaded region only non-zonal coherent structures emerge, while in the dark shaded region both zonal jets and non-zonal coherent structures emerge. In the region
$\b^* < \b_{\min}^*$ (the shaded region to the left of the dashed line) zonal jets have larger growth rate, while in  $\b^*> \b_{\min}^*$ non-zonal structures have the larger
growth rate.  Parameter values for  the Earth's atmosphere, Earth's ocean and Jupiter's atmosphere are marked with stars.}\label{fig:emin}
\end{figure}

The non-zonal structures were treated in these studies as incoherent due to the assumption that
the ensemble average over the forcing realizations is equivalent to a zonal average and therefore
their emergence and effect on the jet dynamics could not be directly addressed. By making the
alternative interpretation of the ensemble mean as a Reynolds average over the fast turbulent
motions that was introduced in earlier studies of atmospheric blocking
\citep{Bernstein-2009,Bernstein-Farrell-2010}, \cite{Bakas-Ioannou-2013-prl,Bakas-Ioannou-2014-jfm}
addressed the emergence of the non-zonal coherent structures in barotropic beta-plane turbulence
in terms of the parameters $\b^* = \b / (r L_f^{-1})$ and
$\e^* = \e/(r^3 L_f^2)$, where $\b$ is the gradient of
the planetary vorticity, $L_f$ the length-scale of the forcing,
$\e$ the energy input rate of the forcing and $1/r$
the dissipation time-scale.  Characteristic  values  of these parameters for the Earth's midlatitude atmosphere and oceans
and the Jovian atmosphere are given in Table~\ref{tab:pla_values}. It was found that for isotropic
forcing the homogeneous statistical equilibrium becomes unstable when
the energy input rate exceeds a critical value $\e^*_c$ that
depends on $\b^*$ as shown in the stability regime diagram in
Fig.~\ref{fig:emin}. In marginally unstable flows with $\b^* \ll 1$ zonal jets
first emerge, while for $\b^*\gg1$ westward propagating non-zonal structures first emerge and equilibrate
to finite-amplitude traveling waves. At larger
energy input rates, the finite-amplitude non-zonal traveling states are
unstable and the flow equilibrates to mixed zonal jet--traveling wave
states that consist of strong zonal jets with weaker traveling non-zonal
structures embedded in them. These predictions of the S3T stability
analysis were verified by direct numerical simulations of  turbulent
barotropic flow~\citep{Bakas-Ioannou-2014-jfm}.

\begin{table*}[t]
\caption{Typical parameter values for geophysical flows. The typical forcing length scale is taken as the deformation radius in each geophysical setting. }\label{tab:pla_values}
\begin{center}
\begin{tabular}{l c c c c c c c}
\topline
 &  $1/k_f$~$[\textrm{km}]$   &   $1/r$~$[\textrm{day}(=24\textrm{h})]$ &   $U_{\textrm{rms}}$~$[\textrm{m}\,\textrm{s}^{-1}]$  & $\b~[10^{-11}\textrm{m}^{-1}\,\textrm{s}^{-1}]$ & $\e~[\textrm{m}^{-2}\,\textrm{s}^{-3}]$ & $\b^*$ & $\e^*$\\
\midline
Earth's atmosphere & 1000 & 10 & $15$ & $1.6$ & $2\times 10^{-3}$ & 15 & 1300\\
Earth's ocean & 20 & 100 & $0.1$ & $1.6$& $10^{-9}$ & 3 & 1600\\
Jovian atmosphere & 1000 & 1500 & $50$ & $0.35$& $0.5\times 10^{-5}$ & 450 & $4\times10^7$\\
\botline
\end{tabular}
\end{center}
\end{table*}

The S3T dynamics that underlie the formation of large-scale structure
cannot depend on turbulent anisotropic inverse cascade processes
because local in wavenumber space eddy--eddy interactions are absent in
S3T. In S3T, large-scale structure emerges from a cooperative
instability arising from the non-local in wavenumber space interaction
between the large-scale mean flow and the forced, small-scale turbulent
eddies. The eddy--mean flow dynamics of this cooperative instability has
been investigated by~\citet{Bakas-Ioannou-2013-jas} for the case of
zonal jet emergence in the limit of ${\b^*}\ll1$. It was shown that
shear straining of the small-scale eddies by the local shear of an
infinitesimal sinusoidal zonal jet, as described by Orr dynamics in a
beta-plane, produces upgradient fluxes that intensify the
zonal jet. In this work we will extend the study of this cooperative
eddy--mean flow instability to address not only zonal jet formation but also
formation of non-zonal coherent structures and also we will address the
formation of coherent structures for a wide range of values
of ${\b^*}$. We will show that for $\b^*\ll1$ the eddy--mean flow dynamics   of non-zonal structures is  also dominated
by shearing of the eddies, whereas for $\b^*\gg1$ resonant and near resonant interactions play an important role in the dynamics. The importance of near-resonant interactions in the formation of large-scale flows has been previously discussed by~\citet{Lee-Smith-2007}. However, this effect is rigorously quantified in this work. Finally, we will discuss the connection between the modulational instability of plane Rossby waves~\citep{Lorenz-1972,Gill-1974} and  the S3T stability of the homogeneous turbulent equilibrium.

This paper is organized as follows: In section~\ref{sec:formulation} we
derive the S3T system for a barotropic flow and the resulting eigenvalue
problem addressing the stability of the homogeneous statistical
equilibrium. In section~\ref{sec:rotation} we transform the eigenvalue
problem in a rotated frame of reference, so that the formation of zonal
jets and non-zonal structures can be studied under a uniform framework.
In section~\ref{sec:iso} we identify the eddy--mean flow dynamics
underlying the S3T instability for isotropic stochastic forcing and in
section~\ref{sec:anisotr} we study the effect of the forcing anisotropy
to the S3T instability. The results are summarized in sections~\ref{sec:discussion} and ~\ref{sec:concl}.

\section{Formulation of S3T dynamics and emergence of non-zonal coherent structures\label{sec:formulation}}

Consider a barotropic flow on an infinite beta-plane with $x$ and $y$ Cartesian coordinates along the zonal and the meridional direction respectively and with planetary vorticity gradient, $\bv=(0,\b)$. The non-divergent velocity field with $(x,y)$
components $(u,v)$  is expressed in terms of the streamfunction, $\psi$, as  $\uv=\zhat\times\nablav\psi$,
where $\zhat$ is the unit vector normal to the plane of the flow. The vorticity of the fluid $\z=\partial_x v-\partial_y u=\Del\psi$, with $\Delta \equiv \partial^2_{xx} + \partial^2_{yy}$, evolves as:
\begin{align}
\partial_t \z + J\(\psi,\z+ \bv\cdot\mathbf{x} \) = -r\z +\sqrt{\e} \,\xi\ ,\label{eq:nl}
\end{align}
where $\mathbf{x}=(x,y)$ and $J$ is the two dimensional Jacobian, $J(A,B)\equiv(\partial_xA)(\partial_yB)-(\partial_yA)(\partial_xB)$. The flow is dissipated with linear damping at a rate $r$, which typically models Ekman drag in planetary atmospheres. Turbulence is maintained by the external stochastic forcing, $\xi$, which models exogenous processes, such as turbulent convection or energy injected by baroclinic instability. We assume that $\sqrt{\e}\,\xi(\xv,t)$ is a temporally delta-correlated and spatially homogeneous random stirring that injects energy at a rate $\e$. We non-dimensionalize (\ref{eq:nl}) using the dissipation time-scale $1/r$ and the typical length-scale of the stochastic excitation, $L_f$. In these units $\zeta^*=\z/r$, $\psi^*=\psi/(rL_f^2)$, $\b^*=\b / (rL_f^{-1})$, $\e^*=\e /(r^3 L_f^2)$, $\xi^*=\xi/(r^{1/2}L_f^{-1})$ and $r^*=1$ where the asterisks denote non-dimensional variables and we hereafter drop the asterisks for simplicity.

\vspace{1em}


In order to construct the S3T dynamical system in the continuous formulation
of \cite{Srinivasan-Young-2012} we proceed as follows:

\begin{enumerate}
\item
The averaged fields, denoted with  upper case letters, are calculated by taking  a time average, denoted
with $\tav{\bullet}$, over an intermediate time scale,  larger than the time scale of
the turbulent motions but smaller than the time scale of the large-scale motions. Deviations from the
mean (eddies) are denoted with dashes and lower case letters. For example the vorticity field
is split as $\z=Z+\z'$, where $Z=\tav{\z}$. The equations for the mean and the eddies that derive
from~\eqref{eq:nl} are:
\begin{subequations}
\begin{align}
\partial_t Z &+ J \(\Psi, Z+ \bv\cdot\mathbf{x} \) = -\tav{ J \(\psi', \z'\)}-Z \ ,\label{eq:enl_mean}\\
\partial_t \z' &=  \Acal(\Uv)\,\z' + f_{\textrm{NL}}+\sqrt{\e}\,\xi \ ,\label{eq:enl_pert}
\end{align}\label{eq:enl}
\end{subequations}
where
\begin{equation}
\Acal(\Uv)\equiv -\Uv\cdot\nablav + \[\bit(\Del\Uv)\cdot\nablav+\zhat\cdot\( \bv\times\nablav\)\]\Del^{-1} -1~,
\label{eq:Aop}
\end{equation}
is the  linear perturbation operator about
the instantaneous mean flow $\Uv=\zhat\times\nablav\Psi$
and  $f_{\textrm{NL}}\equiv\tav{ J \(\psi', \z'\) }-J \(\psi', \z'\)$.
Neglecting the  nonlinear term $f_{\textrm{NL}}$ in \eqref{eq:enl_pert}   we obtain the quasi-linear system:
\begin{subequations}\begin{align}
\partial_t Z &+ J \(\Psi, Z+ \bv\cdot\mathbf{x} \) = -\tav{ J \(\psi', \z'\) }-Z \ ,\label{eq:eql_mean}\\
\partial_t \z' &=\Acal(\Uv)\,\z'+\sqrt{\e}\,\xi \ .\label{eq:eql_pert}
\end{align}\label{eq:eql}\end{subequations}
\item
The quasi-linear system~\eqref{eq:eql} under the ergodic assumption that the time average over
the intermediate time scale is equal to an ensemble average produces the S3T system:
 \begin{subequations}\begin{align}
\partial_t Z &+ J \(\Psi, Z+ \bv\cdot\mathbf{x} \) = \Rcal( C )-\,Z\ ,\label{eq:s3t_mean}\\
\partial_t C & = \[\bit\Acal_a(\Uv) + \Acal_b(\Uv)\]C +\e\,Q \ ,\label{eq:s3t_pert}
\end{align}\label{eq:s3t}\end{subequations}
with $C$ the ensemble mean eddy-vorticity spatial  covariance between points $\xv_a$ and $\xv_b$,
\be
C(\xv_a,\xv_b,t)=\<\z'(\xv_a,t)\z'(\xv_b,t)\>\ ,\label{eq:def_C}
\ee
$Q$ the spatial covariance of the delta-correlated and spatially homogeneous forcing,  defined by
\be
\<\xi(\xv_a,t_1)\xi(\xv_b,t_2)\>  =  Q(\xv_a-\xv_b) \delta(t_1-t_2)\ ,
\ee
and $\Rcal( C ) \equiv \<\,J(\psi',\z')\,\>=\tav{J(\psi',\z')}$  the ensemble mean vorticity forcing
of the large scales by the eddy field, given by:
\begin{align}
\Rcal( C ) \equiv -\nablav\cdot\[ \frac{\zhat}{2}\times(\nablav_a\Del^{-1}_a+\nablav_b\Del^{-1}_b)
C\]_{\xv_a=\xv_b}   \ .\label{eq:defR} 
\end{align}
The subscript $a$ (or $b$) in the operators indicates that  the coefficients of the operator are evaluated on $a$
(or $b$) and that the operator acts only on the variable $\xv_a$ (or $\xv_b$).  The subscript $\xv_a=\xv_b$  indicates
that any function of  $\xv_a$ and $\xv_b$ is evaluated at the same point, $\xv_a=\xv_b$.
\end{enumerate}

The S3T system \eqref{eq:s3t} is a closure of the statistical dynamics of  \eqref{eq:enl}  at
second order.  Being autonomous it may posses statistical equilibria $(Z^e,C^e)$, the stability of
which is addressed by considering small perturbations $(\d Z,\d C)$ and performing an eigenanalysis of the
linearized S3T equations about these equilibria.

\begin{figure*}[t]
\centerline{\includegraphics[width=.65\textwidth,trim = 18mm 8mm 18mm 5mm, clip]{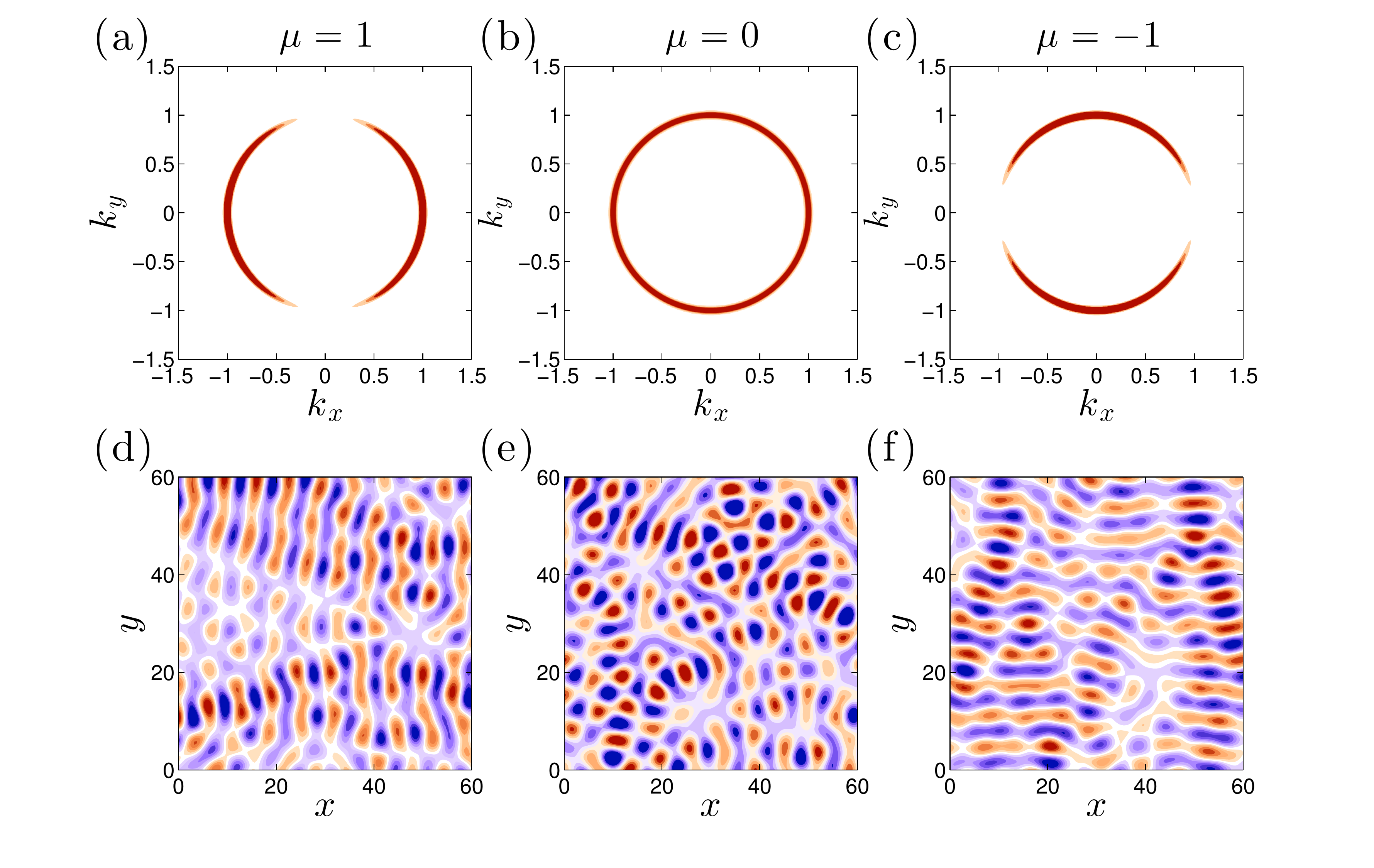}}\vspace{-1em}
\caption{\label{fig:spec_forc} Top panels: the forcing covariance spectrum, $\hat{Q}(\kv)= 4\pi\,\d(k-1)\,\[1+\mu \cos(2\gamma)\]$,  for (a) $\mu=1$, (b) $\mu=0$ and (c) $\mu=-1$ (the support of the delta function is represented as a thin  ring). Bottom panels:  contours of the vorticity field  induced by a realization of the stochastic forcing for (d) $\mu=1$, (e) $\mu=0$ and (f) $\mu=-1$.}
\end{figure*}

For any spatially homogeneous forcing, $Q$, there is always the homogeneous S3T equilibrium
\be
Z^e=0~\ ,~~C^e=\frac{\e}{2} Q\ ,\label{eq:ZeCe}
\ee
with no mean flow and a homogeneous eddy field, i.e., with
a translationally invariant covariance, $C^e(\xv_a-\xv_b)$. The stability of the homogeneous equilibrium~\eqref{eq:ZeCe}
is  determined from eigenalysis of the linearized S3T equations about this equilibrium:
\begin{subequations}\begin{align}
\partial_t \,\d Z &+ J \(\d\Psi, \bv\cdot\mathbf{x} \) = \Rcal( \d C )-\,\d Z\ ,\label{eq:s3t_pert_dZ}\\
\partial_t \,\d C & = \(\bit\Acal^e_a + \Acal^e_b \)\d C +\(\bit\d\Acal_a + \d\Acal_b\)C^e\ ,\label{eq:s3t_pert_dC}
\end{align}\label{eq:s3t_dZdC}\end{subequations}
with
$\Acal^e \equiv\zhat\cdot\( \bv\times\nablav\)\Del^{-1} -1$  (obtained by setting $\Uv^e=0$  in \eqref{eq:Aop})
and $\d\Acal \equiv \Acal(\d\Uv)-\Acal^e$. It can be shown  from~\eqref{eq:s3t_dZdC} that  the homogeneous equilibrium
is S3T stable  for $0\le \e<\e_c$ and becomes unstable when $\e$ exceeds the critical value
$\e_c$ that depends on $\b$ and on the structure of $Q$. In this work we focus our analysis close to
the instability threshold $\e\approx\e_c$ and identify the physical processes
underlying the S3T instability.
We follow~\citet{Srinivasan-Young-2014} and consider  a ring stochastic forcing of waves of total wavenumber $k=|\kv|$=1,
with power spectrum\footnote{The power spectrum of a spatially homogeneous  covariance is  the Fourier transform of the covariance:\bdm\hat{Q}(\kv) =\iintinf Q(\xv_a-\xv_b)\,e^{-\i \kv\cdot(\xv_a-\xv_b)}\df^2 (\xv_a-\xv_b)\ .\edm}:
\be
\hat{Q}(\kv) = 4\pi\,\d(k-1)\,\Gcal(\gamma)\ ,\label{eq:Qhat}
\ee
where
\be
\Gcal (\gamma) = 1 + \mu \cos (2 \gamma)\ ,\label{eq:defG}
\ee
with $\gamma=\arctan{(k_y/k_x)}$ and $\kv=(k_x,k_y)$. The parameter $\mu$  modulates the anisotropy of the spectrum of the forcing and
takes values    $|\mu |\le1$  in order that the spectrum is everywhere positive and therefore physically realizable.
Example  realizations of the spatial structure of stochastic excitations  at different  $\mu$ are shown in Fig.~\ref{fig:spec_forc}.
When $\mu=0$, the forcing is
isotropic and could model  the forcing of the Jovian atmosphere at cloud
level from  turbulent convection. When $\mu>0$, the stochastic
excitation  favors  small $|k_y|$ Fourier components  as the baroclinic forcing of the upper-level
jet in the midlatitude atmosphere. When $\mu<0$, the forcing  favors the almost zonal Fourier components around $k_x=0$.

\section{Emergence of non-zonal structures as zonal flows in a rotated frame\label{sec:rotation} }

The eigenfunctions of the S3T stability equations~\eqref{eq:s3t_dZdC} are specified by their two components: the mean flow component
$\dZ\, e^{\s t}$, and the covariance component $\dC\,e^{\s t}$. Because the stability equations~\eqref{eq:s3t_dZdC} are linearized about the homogeneous equilibrium~\eqref{eq:ZeCe}, the
eigenfunction structure simplifies significantly and assumes the form:
\begin{subequations}\begin{align}
\dZ(\xv )&=e^{\i\nv\cdot\xv}\ ,\label{eq:dZeigen}\\
\dC(\xv_a,\xv_b)&=\tilde C^{(\textrm{h})}_\nv(\xv_a-\xv_b)\,e^{\i\nv\cdot(\xv_a+\xv_b)/2}\ ,
\end{align}\label{eq:s3t_eigenfunctions}\end{subequations}
where $\nv=(n_x,n_y)$ is the wavevector that characterizes the eigenfunction and $\tilde C^{(\textrm{h})}_\nv(\xv_a-\xv_b)$
is the homogeneous component of the covariance eigenfunction. The mean flow component of each
eigenfunction has the form of a zonal flow when $n_x=0$ and of a non-zonal flow when $n_x \ne 0$.
However, non-zonal mean flow perturbations can be rendered zonal through a rotation of the frame
of reference.

\begin{figure}
\centerline{\includegraphics[width=15pc]{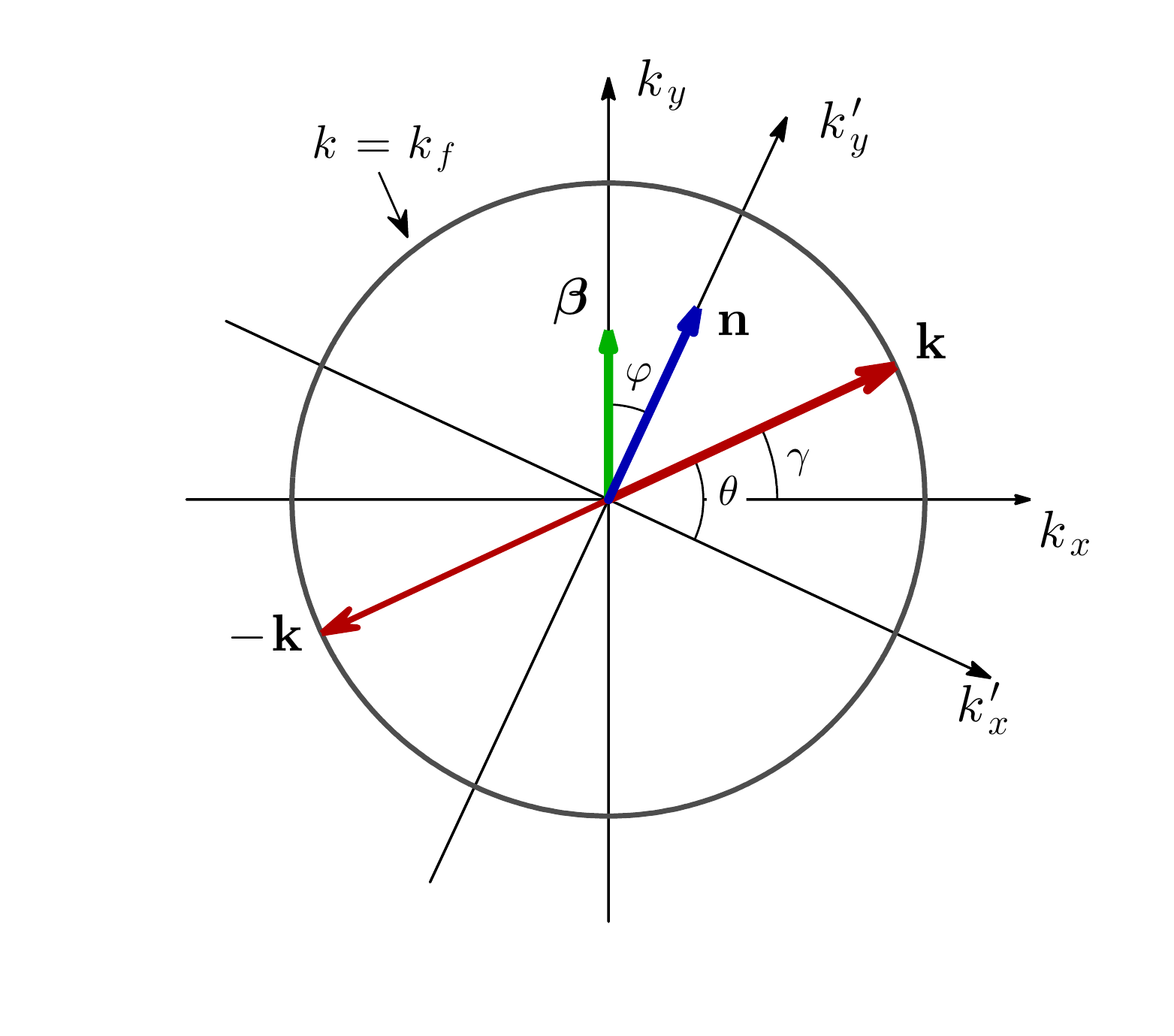}}
\vspace{-2em}
\caption{\label{fig:einx_angles}  A non-zonal plane wave perturbation with   wavevector $\nv$
at an angle  $\varphi$ to the northward direction (the direction of $\bv$)
becomes  a zonal perturbation when the coordinate frame is rotated clockwise by  $\varphi$.  Under this rotation the components of the wavevector  $\kv=(\cos\gamma,\sin\gamma)$ are transformed
to $\kv=(\cos\thet,\sin\thet)$, with $\thet=\gamma+\varphi$.
$\Fcal(n,\thet)$ in~\eqref{eq:fr} is the mean momentum flux convergence from plane wave  perturbations that arise from excitations with wavevectors $\kv$ and $-\kv$.}
\end{figure}

For an eigenfunction with wavenumber $\nv$, clockwise rotation of the axes by an angle $\varphi=\arctan(n_x/n_y)$ transforms the components of $\nv$ to:
\begin{equation}
n_x'= n_x \cos \varphi - n_y \sin \varphi = 0 ~~, ~~~n_y' = n_x \sin \varphi + n_y \cos \varphi = n\ ,
\end{equation}
with $n= |\nv|$ and the components  of the planetary vorticity gradient to $\bv = (-\b\sin\varphi,\b\cos\varphi)$.
Correspondingly, in the rotated frame the eigenfunction has only the mean flow component $\dU(y')$, which is of the form of
 a zonal jet in the $x'$ direction (i.e. the wavevector $\nv$  has zero $x'$ component,
 cf.~Fig.~\ref{fig:einx_angles}), and $\dZ=-\partial_{y'} \dU$. The eigenvalue problem~\eqref{eq:s3t_dZdC} about the
 homogeneous equilibrium transforms to:
\begin{subequations}
\label{eq:dS3TR}
\begin{align}
\sigma \,\partial_{y'}\dU &=-\b\sin\varphi\,\dU -\,\partial_{y'}\dU+ \partial_{y'}\(\bit\dvz\)\ ,\label{eq:dS3TR_dZ}\\
\sigma \, \dC & = \(\bit {\Acal^e_a}' + {\Acal^e_b}' \)\dC +\(\bit\d\Acal'_a + \d\Acal'_b \){C^e}'\ ,\label{eq:dS3TR_dC}
\end{align}
\end{subequations}
with ${\Acal^e}' \equiv -\( \b\sin\varphi \, \partial_{y'} + \b\cos\varphi \, \partial_{x'} \) \Del^{-1} -1$ and $\d\Acal' = -\dU\,\partial_{x'} + (\partial^2_{y'y'}\dU)\,\partial_{x'} \Del^{-1} $. ${C^e}'$ is the equilibrium covariance in the rotated frame defined as ${C^e}'(x_a'-x_b',y_a'-y_b') = {C^e}(x_a-x_b,y_a-y_b)$, where $(x_a', y_a')$ are the components of $\xv_a$ in the rotated frame. The  perturbation
vorticity flux $\dvz$ is given in terms of $\dC$ as:
\be
\dvz = \[ \frac1{2}(\Del^{-1}_a\partial_{x'_a}\!+\!\Del^{-1}_b\partial_{x'_b}) \, \dC\]_{\xv_a=\xv_b}\ .\label{eq:dvz_sl}
\ee
In writing the S3T eigenvalue problem in the rotated frame and by transforming a non-zonal perturbation into a zonal jet perturbation there is a twofold gain. The first is that we can use the methods that were previously developed by \citet{Bakas-Ioannou-2013-jas} in the context of the emergence of zonal jets in order to understand the mechanisms responsible for the emergence of non-zonal structures. The second is that we can directly address
the eddy--mean flow dynamics that give rise to zonal jets with constant topographic vorticity gradient but  in a direction other than the meridional
\citep{Boland-etal-2012}.

From here on we will study  the S3T instability of the homogeneous equilibrium~\eqref{eq:ZeCe}
in  the rotated frame. 
While in the rotated frame  all
eigenfunctions have the form of a zonal jet,  we will distinguish the perturbations as zonal when $\varphi=0\deg$  and  non-zonal  when
$\varphi\neq 0\deg$, i.e., as they manifest in the unrotated frame. A direct implication of~\eqref{eq:dS3TR_dZ} is that
that  the vorticity flux, $\dvz$, induced by eigenfunction $\dU$, $\dC$ must
be proportional to $\dU$ and it  therefore can be written as:
\be
\d\<v'\z'\> = \e\,\f(\s)\,\dU\ ,\label{eq:sens}
\ee
with   $\f$  determining the amplitude and relative phase  of the vorticity flux feedback induced by the mean flow eigenfunction $\dU$
with eigenvalue $\sigma$. When  the real part of  $\f$  is positive  the induced vorticity flux is upgradient, and when it
is  negative the flux is  downgradient.
It is  shown in \hyperref[app:solve_dC]{Appendix~A}  that
\begin{align}
\f(\s) &= \iintinf \frac{\df k'_x\,\df k'_y}{(2\pi)^2} \[ \bit 2n {k'_x}^2 ({k'_y}+n/2)\(k^2 -n^2\)\]\times \nonumber\\
&\qquad\times \[\bit (\sigma+2)k^2_{s}k^2 + 2\i n\b\cos{\varphi}\, {k'_x}  ({k'_y}+n/2) \right.\nonumber\\
&\qquad\left.\bit-  \i n \b \sin{\varphi} \( {k'_x}^2- {k'_y}^2 - n {k'}_y\) \]^{-1}\times  \frac{{\hat{Q}'}({k'}_x,{k'}_y)}{2}~,
\end{align}
with $\kv_s=\kv+\nv$, $k_s=|\kv_s|$ and $(k'_x,k'_y)=k(\cos \thet,\sin \thet)$. The power spectrum of the stochastic forcing, \eqref{eq:Qhat},
in the rotated frame takes the form:
\be
\hat{Q}'(k'_x,k'_y) = 4\pi\,\d(k-1)\,\Gcal(\thet-\varphi)\ .\label{eq:QhatR}
\ee
With this notation the eigenvalues
$\sigma$  of   \eqref{eq:dS3TR}
satisfy the equation:
\begin{align}
 \s+1-\i \beta\sin{\varphi}/n =\e\,\f(\s)\ .\label{eq:S3T_nphi}
\end{align}
For $\varphi=0\deg$,~\eqref{eq:S3T_nphi} reduces
to the eigenvalue relation
of~\citet{Srinivasan-Young-2012} that governs the stability
of the homogeneous equilibrium~\eqref{eq:ZeCe} to zonal jet perturbations in the unrotated frame.
For mirror symmetric  forcing (i.e. with covariance satisfying $Q(x_a-x_b,y_a-y_b) = Q(x_b-x_a,y_a-y_b)$ or $Q(x_a-x_b,y_a-y_b) = Q(x_a-x_b,y_b-y_a)$), like \eqref{eq:Qhat},  and  for  $n<1$ we find numerically\footnote{We have been unable to find a counterexample to these assertions or to prove them when $n<1$. For $n>1$ there exist unstable jet eigenfunctions that have $\s_i\ne 0$ for mirror symmetric forcing.}
that the eigenvalue corresponding to unstable
zonal jet eigenfunctions  have $\sigma_i\equiv\imag (\s) =0$, that is the unstable
zonal jets grow in situ. For $\varphi \ne 0\deg$ the above expression produces the eigenvalue relation
obtained by~\citet{Bakas-Ioannou-2014-jfm} for the growth rate of non-zonal perturbations with
wavenumbers $(n_x,n_y)=(n \sin \varphi, n \cos \varphi)$ in the unrotated frame. The growing
eigenfunctions in this case are numerically found to be propagating ($\sigma_i\neq 0$) and
at marginal stability for $\b\gg1$ their frequency $\sigma_i$ becomes the Rossby wave frequency
\be
\om_\nv \equiv \frac{\zhat\cdot(\bv\times\nv)}{n^2}\ ,\label{eq:def_omRossby}
\ee
for a plane wave with wavevector $\nv$.

From \eqref{eq:S3T_nphi}, we obtain  that a  necessary condition  for S3T instability is
that the real part of the vorticity flux feedback factor, $\real(\f)$, must be  positive. In order to
illuminate the eddy--mean flow dynamics underlying the S3T instability,
we study the behavior of $\real(\f)$ for energy input rates close to $\e_c$.
Near the stability boundary $\sigma_r\equiv\real(\s)\approx 0$ and under the assumption that at marginal
stability\footnote{While the phase speed of the marginally unstable
non-zonal structures almost matches the corresponding Rossby phase speed
for $\b\gg 1$ it overestimates the Rossby phase speed by almost by a
factor of 2 when $\b \sim \Ocal(1)$ or smaller. However, at these values
of $\b$ we have found that the results presented in this work are not
sensitive to the  value of the frequency.} $\sigma_i\approx -\om_\nv = \b
\sin \varphi / n$, the feedback  on the mean flow  for the delta function forcing~\eqref{eq:QhatR} can be written as:
\begin{align}
\fr &\equiv\real\[f(-\i\omega_\nv)\bit\]=\int\limits_{0}^{\pi} \Fcal (\thet, n)\,\df\thet\ ,\label{eq:fr}
\end{align}
where $\Fcal(\thet,n)$  (cf. \hyperref[app:solve_dC]{Appendix~A}) is the contribution to
$\fr$ from Fourier components of the forcing with wavevectors $\kv$ and $-\kv$ (see Fig.~\ref{fig:einx_angles}). When $\fr>0$ the induced vorticity fluxes are upgradient and the critical energy input rate is $\e_c = 1/\fr$.
The integrand
$\Fcal(\thet,n)$ can be alternatively interpreted as the contribution of
the stochastically forced waves or eddies to the vorticity fluxes. These
forced waves have  a total wavenumber
$k=1$ and are characterized only by the angle $\thet$ between their
phase lines and the $y'$ axis. We can isolate the dependence of the
feedback factor on $\b$ by writing $\Fcal(\thet,n) =\bit
F(\thet,n)+F(180\deg+\thet,n)$ with
\be
F(\thet,n)=\frac{\Ncal\,\Dcal_0}{ \Dcal_0^2 + \b^2\,\Db^2 }\ ,\label{eq:defF}
\ee
and, as shown in \hyperref[app:solve_dC]{Appendix~A}, functions $\Ncal$, $\Dcal_0$, $\Db$
independent of $\b$.

In the following sections we will determine the contribution of the various waves
to the vorticity flux feedback and identify the angle $\thet$ that produces the most
significant contribution to this feedback. We will also calculate the feedback factor
$\fr$ as a function of the mean flow wavenumber $n$ for $0 \le \varphi \le 90\deg$. We
will limit our discussion to the emergence of mean flows with $n<1$, i.e., with scale
larger than the scale of the forcing. In section \ref{sec:iso} the analysis is mostly
focused to isotropic forcing ($\Gcal=1$) while the effect of anisotropy is discussed
in section~\ref{sec:anisotr}.

\begin{figure}[h]
\centerline{\includegraphics[width=15pc,trim = 10mm 1mm 10mm 5mm, clip]{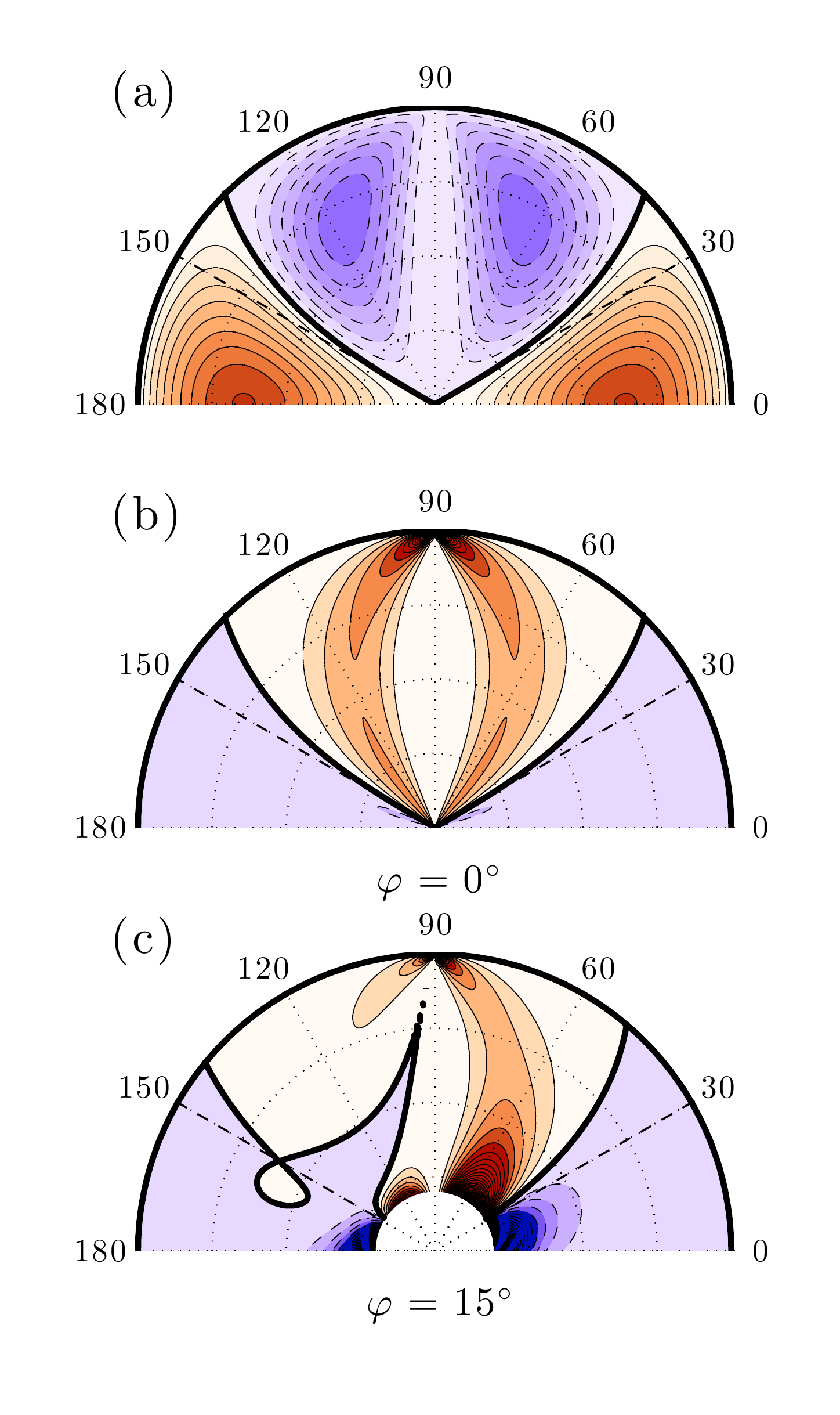}}
\vspace{-2em}
\caption{\label{fig:smallb_F}  (a) Contours of $\Fcal_0(\thet,n)$
in a $(\thet,n)$ polar plot ($n$ radial and $\thet$ azimuthal). This figure shows the magnitude and sign
of the vorticity flux  induced by  waves with phase lines oriented at an angle $\thet$ to the $y$ axis in the presence
of an infinitesimal mean flow perturbation of total wavenumber $n$ when $\b=0$. The contour interval is $3\times 10^{-3}$
and note that $\Fcal_0(\thet,n)$  is independent of $\varphi$. (b) Contours of the normalized $\Fcal_2(\thet,n)/n^4$  show
the $\mathcal{O}(\b^2)$ correction to  $\Fcal_0(\thet,n)$  for the case of zonal jet perturbations ($\varphi=0\deg$). The
contour interval is 0.02.  (c) Same as (b) but for non-zonal perturbations with $\varphi=15^\circ$. The contour interval is
0.04. In all panels the forcing is isotropic ($\mu=0$), solid (dashed) lines indicate contours with positive (negative) values, the thick line  is the zero contour, the radial grid interval is $\Delta n=0.25$ and the $30^\circ$ wedge is marked
(dashed-dot). In panels (a) and (b) the zero contour is the curve $4\sin^2\thet = 1+n^2$ (see~\hyperref[sec:asymptotics]{Appendix~B}).}
\vspace{1em}
\end{figure}

\section{\label{sec:iso}Eddy--mean flow dynamics leading to formation of zonal and
non-zonal structures for isotropic forcing}

\subsection{Induced vorticity fluxes for $\b\ll1$\label{subsec:b0}}

We expand the integrand $\Fcal$ of~\eqref{eq:fr} in powers of
$\b$:
\be
\Fcal=\Fcal_{0} + \b^2\, \Fcal_2 + \Ocal(\b^4)~,
\ee
with $\Fcal_2 = \frac1{2}\left . \bit \partial^2_{\b\b} \Fcal \right|_{\b=0}$.
The leading order term, $\Fcal_{0}$, is the contribution of each wave
with wavevector $\kv=(\cos \thet,\sin \thet)$ to the vorticity flux
feedback in the absence of $\b$ and is shown in
Fig.~\ref{fig:smallb_F}\hyperref[fig:smallb_F]{a}. For $\b=0$, the
dynamics are rotationally symmetric and for isotropic forcing $\fr$ is
independent of $\varphi$. Therefore all zonal and non-zonal
eigenfunctions with the same wavenumber, $n$, grow at the same rate.
Upgradient fluxes ($\Fcal_0>0$) to a mean flow with wavenumber $n$ are
induced by waves with phase lines inclined at angles satisfying
$4\sin^2\thet < 1+n^2$ (cf. \hyperref[sec:asymptotics]{Appendix~B}).
This implies that all waves with $|\thet| < 30 \deg$ necessarily produce
upgradient vorticity fluxes to any mean flow with wavenumber $n<1$,
while waves with $30\deg<|\thet|<45\deg$ produce upgradient fluxes for any mean
flow with large enough wavenumber (cf.~Fig.~\ref{fig:smallb_F}\hyperref[fig:smallb_F]{a}). The eddy--mean flow
dynamics was investigated in the limit of $n\ll1$
by~\citet{Bakas-Ioannou-2013-jas}. It was shown that the vorticity
fluxes can be calculated from time averaging the fluxes over the life
cycle of an ensemble of localized stochastically forced wavepackets
initially located at different latitudes. For $n\ll 1$, the wavepackets evolve
in the region of their excitation under the influence of the
infinitesimal local shear of $\delta U$ and are rapidly dissipated
before they shear over. As a result, their effect on the mean flow is
dictated by the instantaneous (with respect to the shear time scale)
change in their momentum fluxes. Any pair of wavepackets having a
central wavevector with phase lines forming angles $|\thet|<30\deg$ with
 the $y$ axis surrender instantaneously momentum to the mean flow and
reinforce it, whereas pairs with $|\thet|>30\deg$ gain instantaneously
momentum from the mean flow and oppose jet formation. Therefore,
anisotropic forcing that injects significant power into Fourier
components with $|\thet|<30\deg$ (such as the forcing from baroclinic
instability that primarily excites Fourier components with $\thet=0\deg$)
produces robustly upgradient fluxes that asymptotically behave
anti-diffusively. That is, for a sinusoidal mean flow perturbation $\dU = \sin{( n y
)}$ we have $\int_0^\pi\Fcal_0\,\df\thet = K n^2 $ with
$K$ positive and proportional to the anisotropy factor $\mu$
(cf.~\hyperref[sec:asymptotics]{Appendix~B} and \citet{Bakas-Ioannou-2013-jas}).

\begin{figure}
\centerline{\includegraphics[width=17pc]{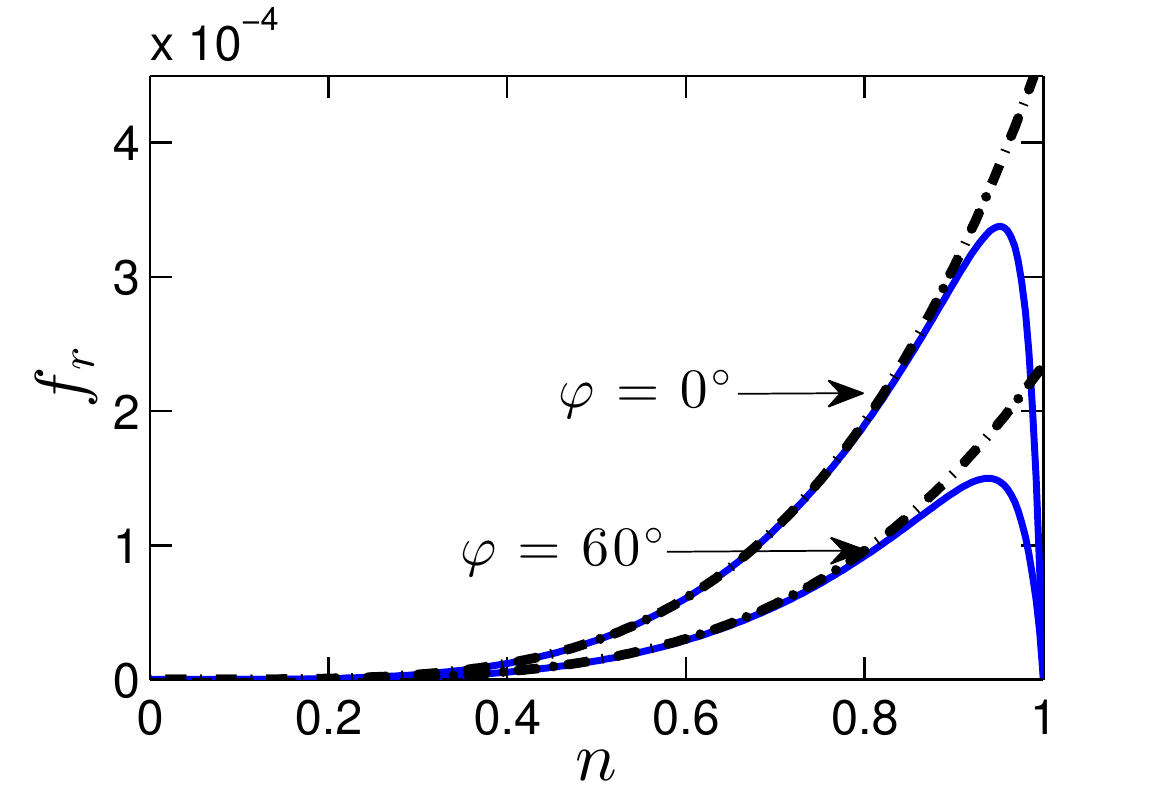}}
\caption{\label{fig:vq_n_phi0_phi60_b0p1} Vorticity flux feedback factor $\fr$ as a function
of $n$ for $\b=0.1$ and isotropic forcing. Note that the fluxes are upgradient (i.e $\fr>0$) for all mean
flow wavenumbers $n$. Shown is $\fr$ for $\varphi=0\deg$ and $\varphi=60^\circ$ (solid lines), as well as the
asymptotic expression~\eqref{eq:fr_bll1_mu0} (dash-dot) derived for the feedback factor in the limit $\b\ll 1$
and $\b/n\ll 1$.}
\end{figure}

For isotropic forcing the net vorticity flux produced by shearing of
the perturbations vanishes, i.e., $\int_0^\pi\Fcal_0\,\df\thet=0$, given
that the upgradient fluxes produced by waves with $|\thet|<30\deg$
exactly balance the downgradient fluxes produced by the waves with
$|\thet|>30\deg$. However, a net vorticity flux feedback is produced
and asymptotically behaves as a negative fourth order hyperdiffusion
with coefficient $\Ocal(\b^2)$ for $\b\ll 1$
(cf.~(\ref{eq:fr_bll1_mu0}) and \citet{Bakas-Ioannou-2013-jas}). In
\hyperref[sec:asymptotics]{Appendix~B} it is shown that the feedback
factor $\fr$ for isotropic forcing in the limit $\b\ll1$ with
$\b/n\ll1$ is: \begin{align} \fr &= \b^2 \frac{n^4}{64}
\[\bit2+\cos(2\varphi)\]+\Ocal(\b^4)\ ,\label{eq:fr_bll1_mu0} \end{align}
which is accurate even up to $n\approx 1$, as shown in
Fig.~\ref{fig:vq_n_phi0_phi60_b0p1}.
In order to understand the contribution of $\beta$ to the vorticity
flux feedback, we plot $\Fcal_2 / n^4 $ for a zonal
(Fig.~\ref{fig:smallb_F}\hyperref[fig:smallb_F]{b}) and a non-zonal
perturbation (Fig.~\ref{fig:smallb_F}\hyperref[fig:smallb_F]{c}) as a
function of the mean flow wavenumber $n$ and wave angle $\thet$. We
choose to scale $\Fcal_2$ by $n^4$ because in \eqref{eq:fr_bll1_mu0} $f_r$
increases as $n^4$. Consider first the case of a zonal jet. It can be
seen that at every point, $\Fcal_2$ has the opposite sign to $\Fcal_0$,
implying that $\b$ tempers both the upgradient (for roughly $|\thet|<30\deg$)
and the downgradient (for $|\thet|>30\deg$) fluxes of $\Fcal_0$. However, in
the sector $|\thet|>30\deg$ the values of $\Fcal_2$ are much larger than
in the sector $|\thet|<30\deg$ and  the net  fluxes
integrated over all angles  are upgradient, as in~\eqref{eq:fr_bll1_mu0} for  the isotropic case.

The asymptotic analysis of \citet{Bakas-Ioannou-2013-jas}, which is
formally valid for $n\ll1$, offers understanding of the dynamics that
lead to the inequality $\Fcal_2 \Fcal_0<0$ and to the positive net
contribution of $\Fcal_2$, i.e., to $ \int_0^\pi\Fcal_2 \,\df \thet>0$.
Any pair of wavepackets with wavevectors at angles $|\thet|>30\deg$
instantaneously gain momentum from the mean flow as described above
(i.e. $\Fcal_0<0$ for $|\thet|>30\deg$), but their group velocity is
also increased (decreased) while propagating northward (southward). This 
occurs due to the fact that shearing changes their meridional wavenumber 
and consequently their group velocity. The instantaneous
change in the momentum fluxes resulting from this speed up (slowing down) 
of the wavepackets is positive in the region of excitation leading to
upgradient fluxes ($\Fcal_2>0$). The opposite happens for pairs with
$|\thet|<30\deg$ (cf. Fig. 3 in \citet{Bakas-Ioannou-2013-jas}), however the
downgradient fluxes produced are smaller
than the upgradient fluxes, leading to a net positive contribution when
integrated over all angles. Figure~\ref{fig:smallb_F}\hyperref[fig:smallb_F]{b}, shows
that this result is valid for larger mean flow wavenumbers as well.

Consider now the case of a non-zonal perturbation (Fig.~\ref{fig:smallb_F}\hyperref[fig:smallb_F]{c}). We observe that the
angles for which the waves have significant positive or negative
contributions to the vorticity flux feedback are roughly the same as in
the case of zonal jets. In addition, the vorticity flux feedback factor
decreases with the angle $\varphi$ of the non-zonal
 perturbations (cf.~\eqref{eq:fr_bll1_mu0}). As a result, zonal jet perturbations
always produce larger vorticity fluxes compared to non-zonal
perturbations and are therefore the most unstable in the limit $\b\ll1$.
Additionally, these results show that for $\b\ll1$, the mechanism for
structural instability of the non-zonal structures is the same as the
mechanism for zonal jet formation, which is shearing of the eddies by
the infinitesimal mean flow.

\begin{figure}
\centerline{\includegraphics[width=19pc]{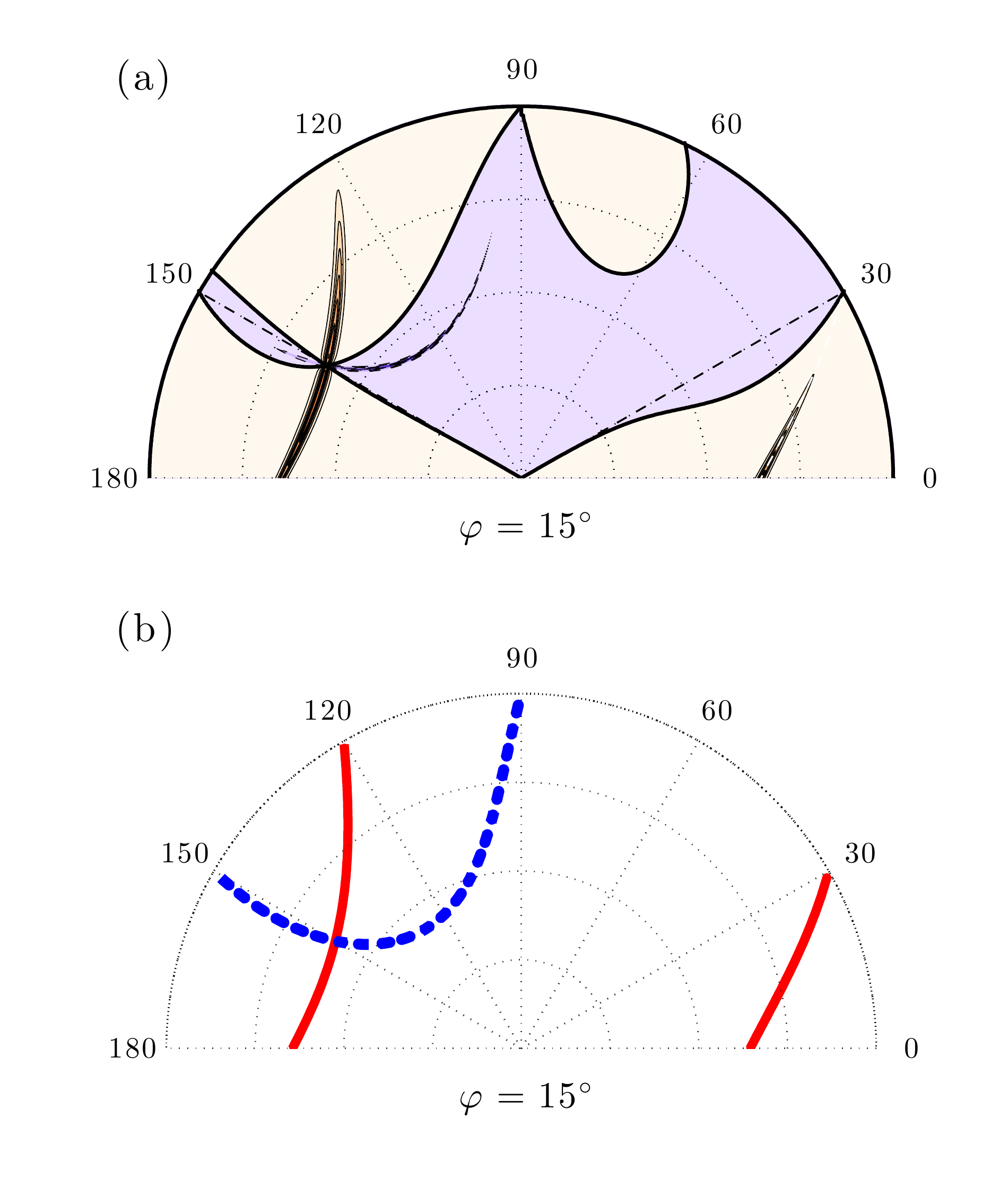}}
\caption{\label{fig:Fr_Db} (a) Contours of $\Fcal(\thet,n)$ in a $(\thet,n)$ polar plot
($n$ radial and $\thet$ azimuthal) for isotropic forcing and $\b=200$. This panel shows the
vorticity fluxes induced  by waves with phase lines oriented at an angle $\thet$ to the $y'$ axis
in the presence of a non-zonal perturbation with  mean flow wavenumber $n$ and $\varphi=15^\circ$.
Solid (dashed) lines indicate contours with positive (negative) values, the contour interval is
$2.5\times10^{-3}$ and the thick line is the zero contour. (b) Locus of the roots of  $\Db(\thet,n)$ on the $(\thet,n)$ plane
for non-zonal perturbations with $\varphi=15\deg$. The roots correspond to resonant interaction  between
waves with phase lines oriented at
an angle $\thet$ with the $y'$ axis and non-zonal perturbations with mean flow
wavenumber $n$.
Thick solid (dashed) lines indicate whether the vorticity fluxes produced by the resonant waves are upgradient (downgradient). The radial grid interval in both panels is $\Delta n=0.25$.}
\end{figure}

\begin{figure*}[t]
\centerline{\includegraphics[width=.8\textwidth]{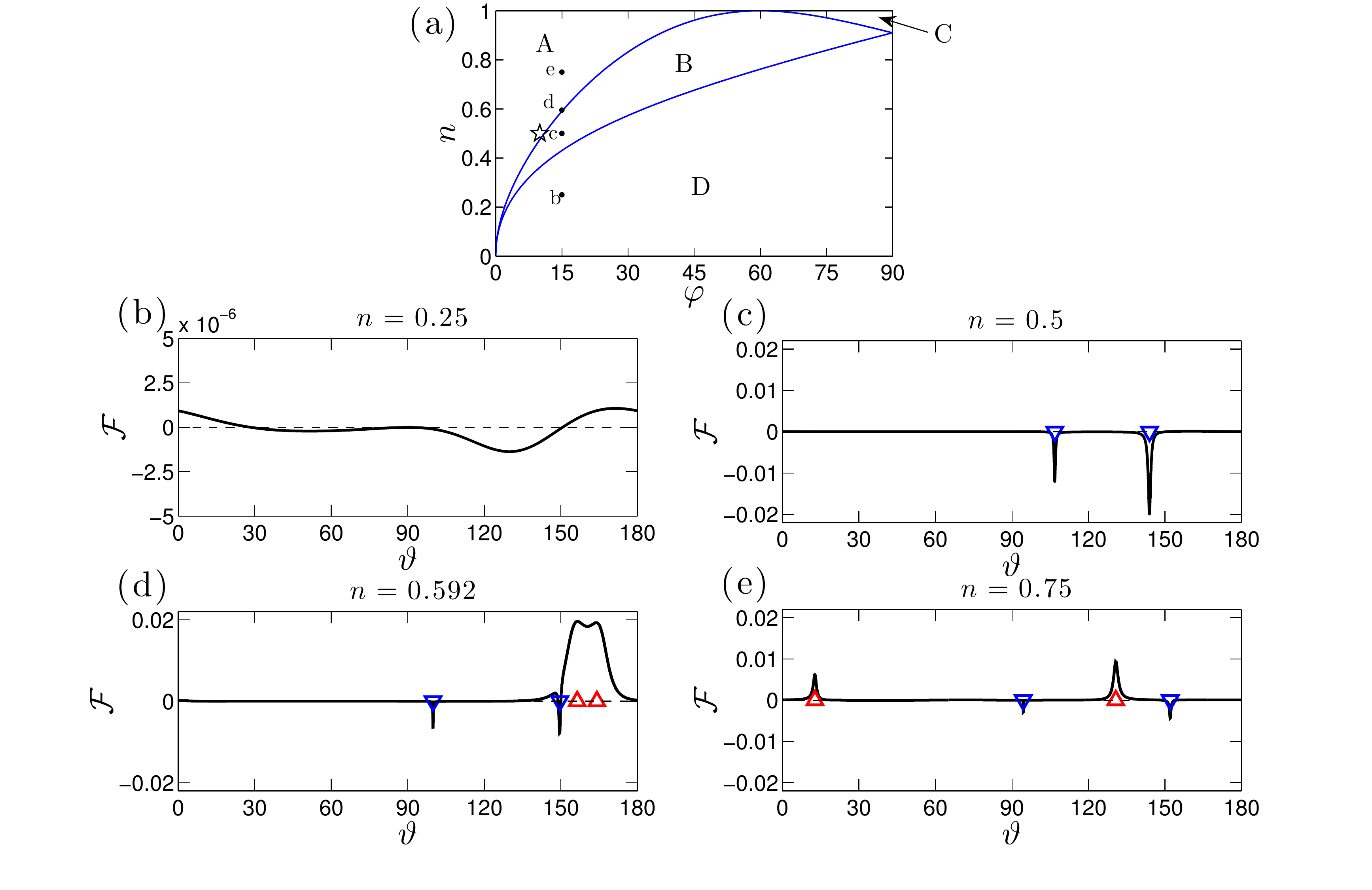}}
\caption{\label{fig:R_Fcal} (a) The curves separating the regions in the $(n,\varphi)$ plane for which $\Db$ has no roots
(region~D), 2 roots (region~B) and four roots (regions A and C). Waves with $\thet$ corresponding to two out of the four
roots of $\Db$ found in region~A produce upgradient fluxes. (b)-(d) The vorticity fluxes $\FcalR$ as a function of the
angle $\thet$ subtended by the phase lines of the waves and the $y$ axis in the presence of a non-zonal perturbation
with $\varphi=15^\circ$ at $\b=200$. The mean flow wavenumber is (b) $n=0.25$ (in region~D), (c) $n=0.5$ (in region~B),
(d) $n=0.592$ (in region~A) and (e) $n=0.75$ (in region~A). The resonant angles (i.e. the roots of $\Db$) are marked by
upper (lower) triangles when the waves induce upgradient (downgradient) fluxes. Note that the scale in (b) is much smaller.}
\end{figure*}

\subsection{Induced vorticity fluxes  for $\b\gg1$ \label{sec:largeb_limit}}

Consider first the emergence of non-zonal structures in the limit
$\b\gg1$. The contribution of each Fourier component of the forcing to
the vorticity flux feedback $\Fcal$ for the case of non-zonal structures
at $\b = 200$ is shown in Fig.~\ref{fig:Fr_Db}\hyperref[fig:Fr_Db]{a}.
In contrast to the cases with $\b\ll1$ (or $\b=\Ocal(1)$, discussed in
section~\ref{sec:iso}\ref{subsec:O1}), there is only a small band of
Fourier components that contribute significantly to the vorticity flux
feedback, as indicated with the narrow tongues in
Fig.~\ref{fig:Fr_Db}\hyperref[fig:Fr_Db]{a}. The reason for this
selectivity in the response is that for $\b\gg1$ the components that
produce appreciable fluxes, as seen from~\eqref{eq:defF}, are
concentrated on the $(\thet,n)$ curves that satisfy $\Db=0$ (shown in
Fig.~\ref{fig:Fr_Db}\hyperref[fig:Fr_Db]{b}) or equivalently for the
$(\thet,n)$ that satisfy the resonant condition
$\om_\kv+\om_\nv=\om_{\kv+\nv}$ (cf.~\eqref{eq:D2zero}). This is the
resonant condition satisfied when a Rossby wave with wavevector $\kv$
and frequency $\om_\kv$ forms a resonant triad with the non-zonal
structure with wavevector $\nv$ and frequency $\om_\nv$.  We concentrate
our analysis to these `resonant contributions'  because  they  dominate the vorticity flux feedback of non-zonal perturbations for $\b \gg 1$.

Resonant triads do not occur for all mean flow perturbations $\nv$. For
$(n, \varphi)$ in region~D of
Fig.~\ref{fig:R_Fcal}\hyperref[fig:R_Fcal]{a}, $\Db$ has no roots and
therefore there are no Fourier components with
$\kv=(\cos\thet,\sin\thet)$ that form a resonant triad with the mean
flow perturbation $\nv$ and the vorticity flux feedback is determined
by the sum over the non-resonant contributions as illustrated in
Fig.~\ref{fig:R_Fcal}\hyperref[fig:R_Fcal]{b}. In region~B of
Fig.~\ref{fig:R_Fcal}\hyperref[fig:R_Fcal]{a}, there are only two
resonant angles $\thet$. The resonant and non-resonant contribution for
a typical case in region~B is shown in
Fig.~\ref{fig:R_Fcal}\hyperref[fig:R_Fcal]{c}. Note that it is the
resonant contributions that determine the vorticity flux feedback.
However, they produce a negative vorticity flux feedback (a
downgradient tendency), which is stabilizing, a result that holds for
all $(n,\varphi)$ in region~B. In regions A and C, there exist four
resonant angles $\thet$ which dominate the vorticity flux. In C all
resonant contributions are stabilizing and therefore C is also a stable
region. In region~A, which at most extends to $\varphi=60\deg$ (cf.~\hyperref[sec:asymptotics]{Appendix~B}),
two of the four resonances give positive contributions to $\fr$
(cf.~Figs.~\ref{fig:R_Fcal}\hyperref[fig:R_Fcal]{d,e}). Therefore only
for $(n,\varphi)$ in region~A, does a destabilizing vorticity flux
feedback occur. The largest destabilizing feedback occurs when the
positively contributing resonances are near coalescence (i.e.~as in
Fig.~\ref{fig:R_Fcal}\hyperref[fig:R_Fcal]{d}), which occurs for $(n,\varphi)$ close to the
curve separating regions A and B. The reason is that when the
resonances are apart, as in Figs.~\ref{fig:R_Fcal}\hyperref[fig:R_Fcal]{c,e}, the
significant contributions come from near-resonant waves with angles within a
band of $\Ocal(1/\b)$ around the resonant angles and the integrated resonant
contributions to the vorticity flux are $\Ocal(1/\b)$. However, when the
resonances are near coalescence, as for the case shown in Fig.~\ref{fig:R_Fcal}\hyperref[fig:R_Fcal]{d},  the band of near-resonant waves contributing significantly increases
as the integrand assumes a double humped shape and, as shown in \hyperref[sec:asymptotics]{Appendix~B}, the
destabilizing vorticity flux feedback becomes $\Ocal(1/\sqrt{\b})$.
Note that as $\b\to\infty$, the width over which we have significant contributions
diminishes and therefore $\fr\to0$ unless an infinite amount of energy is injected
exactly at the resonant angles (as is assumed in modulational instability studies).

\begin{figure}
\centerline{\includegraphics[width=17pc]{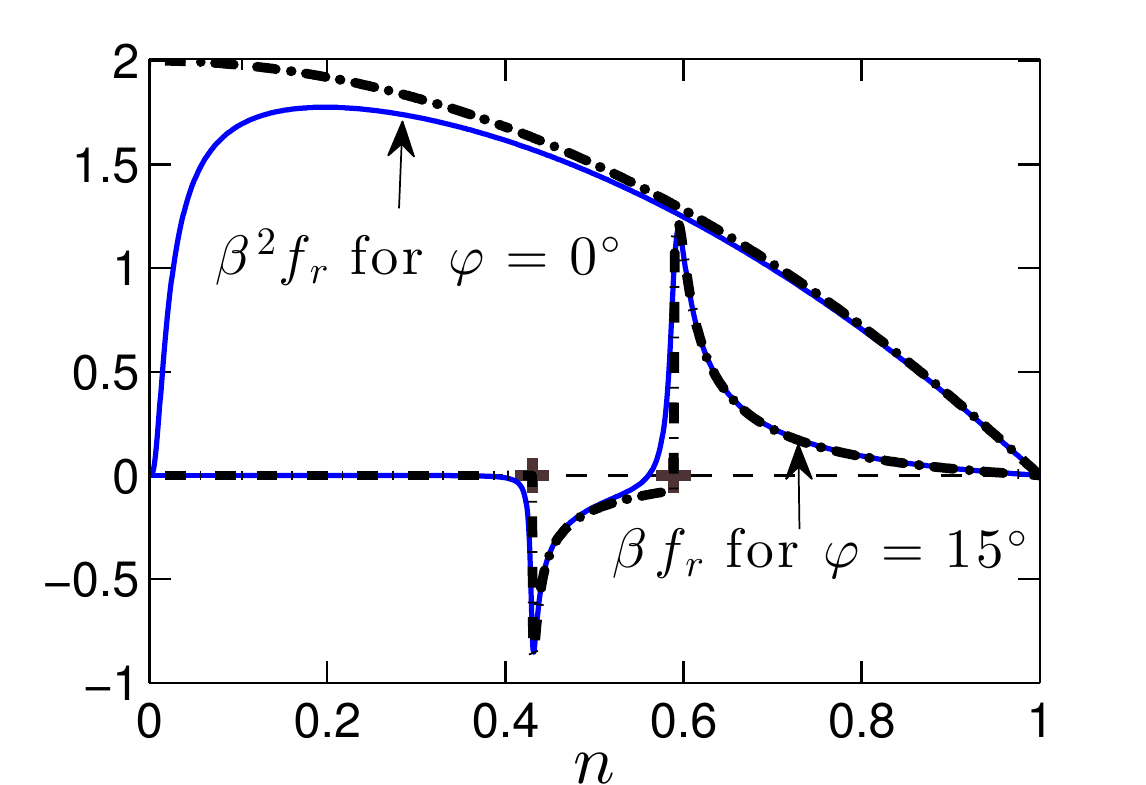}}
\caption{\label{fig:fr_n_b200} Vorticity flux feedback $\fr$ as a function of $n$ for $\b=200$. Positive
(negative) values correspond to upgradient (downgradient) fluxes. Shown is $\fr$ for $\varphi=0\deg$
(multiplied by $\beta^2$) and for $\varphi=15^\circ$ (multiplied by $\beta$). Also shown are the asymptotic
expressions~\eqref{eq:dvzR_largeb} for $\varphi =0\deg$ and~\eqref{eq:frR} for $\varphi=15\deg$ (dash-dot). The
crosses mark the mean flow wavenumbers $n=0.43$ and $n=0.59$ that separate regions A, B and D in
Fig.~\ref{fig:R_Fcal}\hyperref[fig:R_Fcal]{a} for $\varphi=15^\circ$.}
\end{figure}

It can be shown (cf.~\hyperref[sec:asymptotics]{Appendix~B}) that the resonant contribution for $\b\gg 1$ asymptotically approaches
\be
\fr^{(\textrm{R})} = \frac{1}{\sqrt{\b}}\sum_{j=1}^{N_r}\frac{\pi \, \Ncal_j\,\eta_j}{ 2\,\Dcal_{0,j}^{1/2}|\la_j|^{1/2}}\ ,\label{eq:frR}
\ee
where the subscript $j$ 
functions at the $j$-th out of the $N_r$ roots of $\Db$ and $\lambda =
\partial^2_{\thet\thet}\Db$. The values $\Ncal_j$, $\Dcal_{0,j}$,
$\la_j$ are all $\Ocal(1)$, whereas $\eta_j$ is always positive and the
only quantity that has dependence on $\b$. It is $\Ocal(1)$ only for $(n,\varphi)$
just above the separating boundaries of regions A and B and regions B and D in
Fig.~\ref{fig:R_Fcal}\hyperref[fig:R_Fcal]{a} yielding
$\fr^{(\textrm{R})}\sim 1/\sqrt{\b}$ and is $\Ocal(1/\sqrt{\b})$
elsewhere yielding $\fr^{(\textrm{R})}\sim 1/\b$, as also qualitatively
described above. The sign of the $j$-th resonant contribution to the
total vorticity flux feedback
depends only on the sign of $\Ncal_j$. For $(n,\varphi)$ just above the
boundary separating regions B and D, $\Ncal_j<0$ and $\fr$ attains its
minimum value, which corresponds to the largest stabilizing tendency.
This is illustrated in Fig.~\ref{fig:fr_n_b200}, showing the flux
feedback $\fr$ as a function of $n$. For $(n,\varphi)$ just above the
boundary separating regions A and B, coalescence of the two positive
contributing resonances occurs and $\fr$ attains its maximum value,
which corresponds to the largest destabilizing tendency. For small mean
flow wavenumbers $n$ (corresponding to region~D) the feedback factor is
negative and $\Ocal(\b^{-2})$ due to the absence of resonant
contributions.

\begin{figure}
\centerline{\includegraphics[width=19pc]{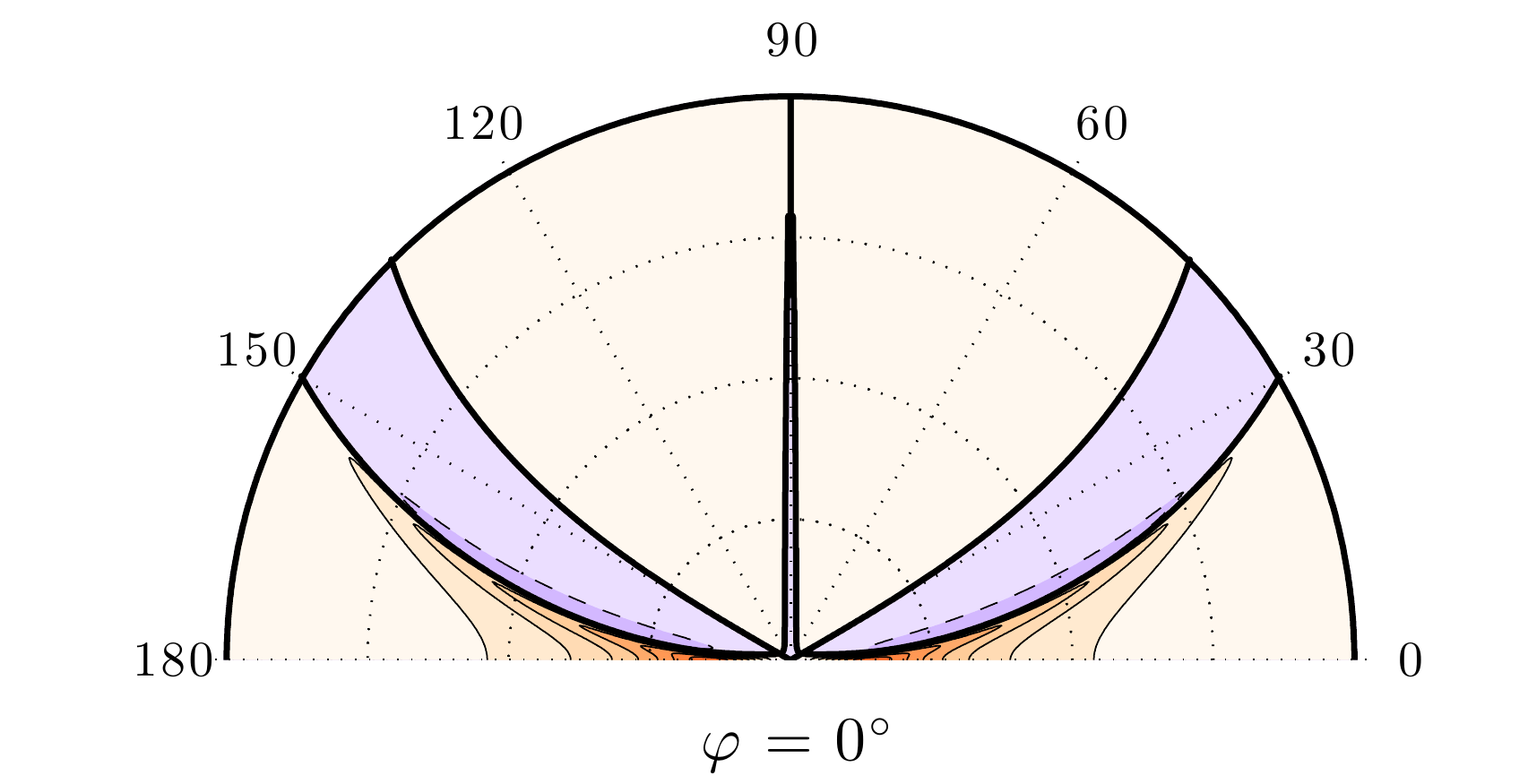}}
\caption{\label{fig:F_z_b200} Contours of $\FcalR(\thet,n)$ in a $(\thet,n)$ polar plot
($n$ radial and $\thet$ azimuthal) for zonal jet perturbations ($\varphi=0\deg$) and $\b=100$. Solid (dashed) lines
indicate contours with positive (negative) values, the contour interval is $2\times10^{-4}$, the thick line
is the zero contour and the radial grid interval is $\Delta n=0.25$.}
\end{figure}

An interesting exception to the results discussed above occurs for the
important case of zonal jet perturbations ($\varphi=0\deg$). In that case,
$\Ncal_j=0$ in~\eqref{eq:frR} as the roots of $\Db$ and $\Ncal$ coincide
and the resonant contribution \eqref{eq:frR} is exactly zero. As shown
in Fig.~\ref{fig:F_z_b200}, positive vorticity flux feedback is obtained from a
broad band of the non-resonant Fourier components with $\gamma=\thet \approx
0\deg$, corresponding to waves with lines of constant phase nearly
aligned with the $y$ axis (remember that for smaller $\b$ the region
that produces destabilizing fluxes extends up to $|\thet| \approx
30\deg$). For large $\b$ the vorticity flux $\fr$ is always
destabilizing for all zonal jet perturbations with $n<1$, as shown by
\eqref{eq:dvzR_largeb} and Fig.~\ref{fig:fr_n_b200}, and the largest
destabilizing vorticity flux, $\f_{r,\textrm{max}}=(2+\mu) \b^{-2}$, is
obtained for jets with the largest allowed scale. The reason for the
weak fluxes and the preference for the emergence of jets of the largest
scale in this limit is understood by noting that the stochastically
forced eddies for $\b\gg1$ propagate with $\Ocal(\b)$ group
velocities. Therefore in contrast to the limit of $\b\ll 1$ in which
they evolve according to their local shear, the forced waves will
respond to the integrated shear of the sinusoidal perturbation over
their large propagation extend, which will be very weak.

\begin{figure}
\centerline{\includegraphics[width=17pc]{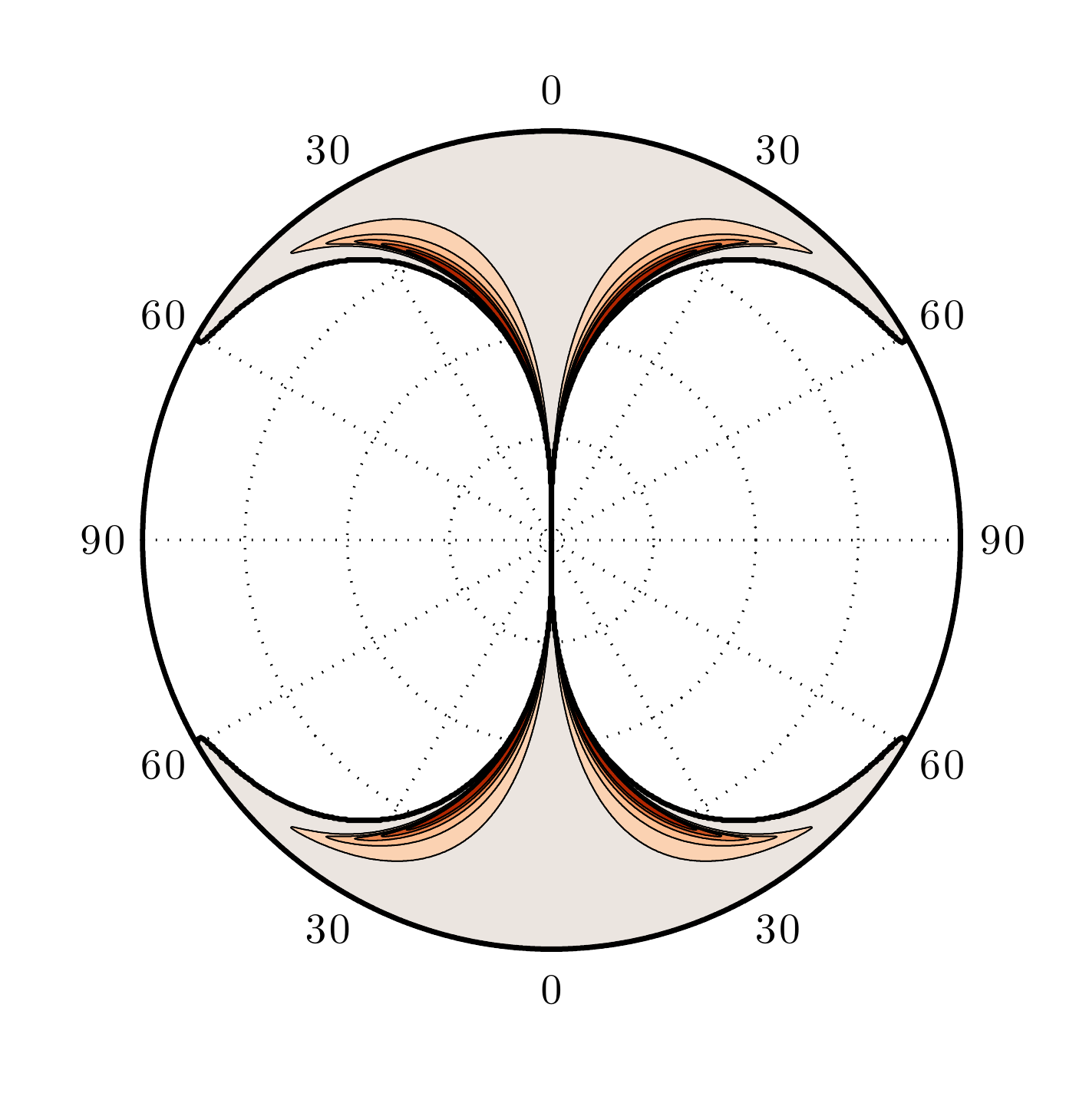}}\vspace{-2em}
\caption{\label{fig:dumbbell_b100}  Contours of the feedback factor $\fr$ in a $(\varphi,n)$ polar plot
($n$ radial and $\varphi$ azimuthal) for the case $\b=200$. Shown are contours of positive values, so the white area
corresponds to negative values indicating downgradient vorticity fluxes. The contour interval is $10^{-3}$ and the radial
grid interval is $\Delta n=0.25$. Note that the feedback factor is always negative (downgradient fluxes) for
$\varphi\ge 60\deg$ (cf. \hyperref[sec:asymptotics]{Appendix~B}).}
\end{figure}

To summarize: Although zonal jets and most non-zonal perturbations
induce fluxes that decay as $1/\b^2$ for large $\b$, resonant
and near resonant interactions arrest the decay rate of certain non-zonal
perturbations by a factor of $\Ocal(\b^{3/2})$ leading to fluxes that decay as
$1/\sqrt{\b}$. This makes the non-zonal perturbations to be the most S3T
unstable perturbations for $\b \gg 1$. Also in contrast to $\b\ll1$
when $\fr$ is positive for all $n$ and $\varphi$
(cf.~Fig.~\ref{fig:vq_n_phi0_phi60_b0p1}), the vorticity flux feedback
is negative for $(n,\varphi)$ in regions~B and D of
Fig.~\ref{fig:R_Fcal}\hyperref[fig:R_Fcal]{a}. As a result, the mean
flows that produce negative fluxes and are by necessity S3T stable are
interestingly in the interior of the dumbbell shown in
Fig.~\ref{fig:dumbbell_b100}. The largest
destabilizing fluxes occur in the narrow region adjacent to the outer
boundaries of the dumbbell shape, which demarcates the boundary
separating regions A and B. Because of the selectivity of the resonances
these results do not depend on the forcing anisotropy as we will see in the
next section.

\subsection{Induced vorticity fluxes for $\b\sim\Ocal(1)$}
\label{subsec:O1}

\begin{figure*}[t]
\centerline{\includegraphics[width=.7\textwidth]{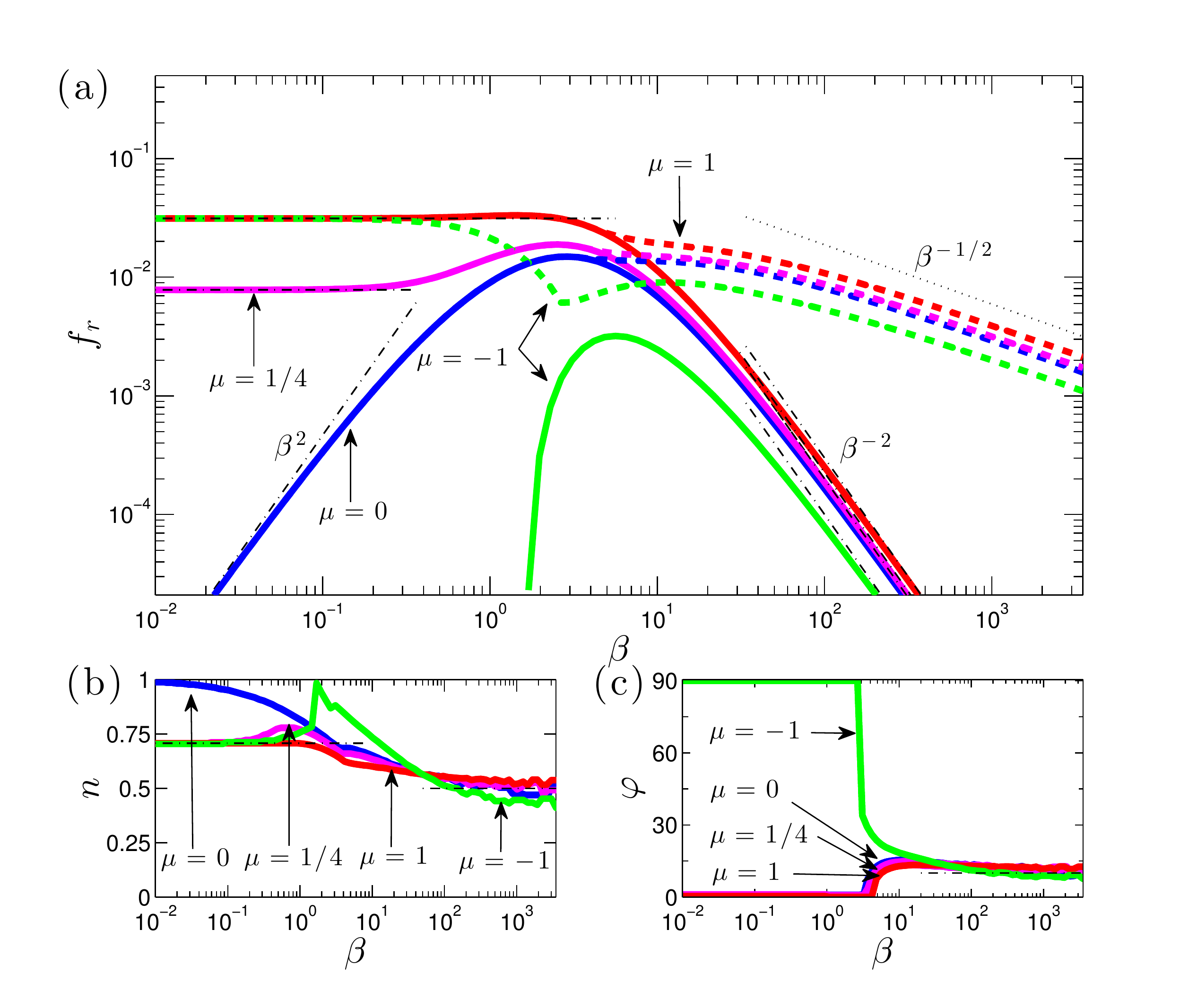}}
\caption{\label{fig:frmax} The maximum value of $\fr$ over all wavenumbers $n$ for zonal jets
(solid), and the maximum value of $\fr$ over all wavenumbers $n$ and angles $\varphi\ne0\deg$ for non-zonal perturbations
(dashed) as a function of the planetary vorticity, ${\beta}$ for the three forcing covariance spectra seen in Fig.~\ref{fig:spec_forc} and for $\mu=1/4$. Also shown are the asymptotic expressions~\eqref{eq:frmax_smallb_mu},
\eqref{eq:frmax_smallb_mu0} and~\eqref{eq:sr_largeb_max} (dash-dot) and the $\b^{-1/2}$ slope (dotted).
For $\mu=-1$ zonal jet perturbations are stable for $\b<1.67$. (b) The mean flow wavenumber $n$ and (c) the angle $\varphi$
for which the maximum value of $\fr$ (shown in (a)) is attained. The asymptotes $n=1/\sqrt{2}$ (for $\b\ll1$) and $n=0.5$
(for $\b\gg1$) are shown in (b) (dash-dot) as well as the asymptote $\varphi=10\deg$ (for $\b\gg1$) is also shown in (c)
(dash-dot).}
\vspace{1em}
\centerline{\includegraphics[width=.69\textwidth,trim = 30mm 1mm 30mm 1mm, clip]{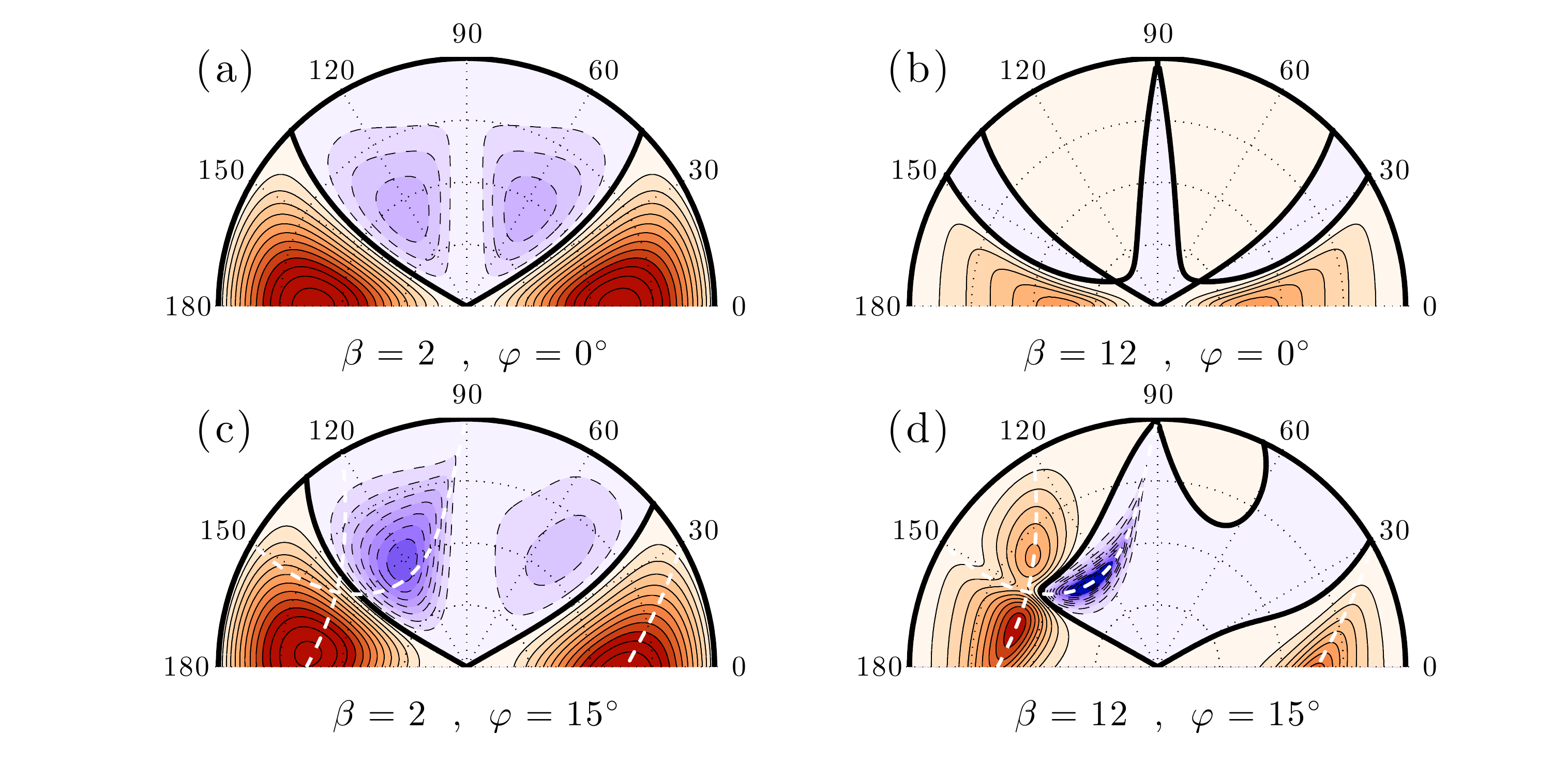}}
\caption{\label{fig:F_order1}  Contours of the $\Fcal(\thet,n)$ in a $(\thet,n)$ polar plot ($n$ radial and
$\thet$ azimuthal). Shown is (a) $\Fcal$ for a zonal jet perturbation ($\varphi=0\deg$) and (c) a non-zonal perturbation
with $\varphi=15^\circ$ when $\b=2$. Panels (b) and (d) are the same as (a) and (c) for the case $\b=12$. In all panels,
solid (dashed) lines indicate contours with positive (negative) values, the contour interval is $2\times10^{-3}$, the
thick lines indicate the zero contour and the radial grid interval is $\Delta n=0.25$. White dashed lines in (c), (d)
correspond to the locus of the roots of $\Db(\thet,n)$ on the $(\thet,n)$ plane.}
\end{figure*}

We have seen that in the singular case of isotropic forcing the only process available for the
emergence of mean flows is the fourth order antidiffusive vorticity
feedback induced by the variation of the group velocity of the forced
eddies due to the mean flow shear. For $\b\ll1$, the waves interact with
the local shear producing fluxes proportional to $\b^2\, \df ^4 \delta U
/ \df y^4 $. As $\b$ increases this growth is reduced since the waves
interact with an effective integral shear within their propagation
extent which is weak and eventually, as we have seen in the previous
section, for $\b\gg1$ the fluxes decay as $\b^{-2}$. Therefore, the
fluxes attain their maximum at an
intermediate value of $\b$. This occurs for $\b\approx 3.5$, as can be
seen in Fig.~\ref{fig:frmax}\hyperref[fig:frmax]{a} where the maximum
$\fr$ over all $(n,\varphi)$ is shown. It will be demonstrated in the
next section that this intermediate $\b$ maximizes the S3T instability
for all forcing spectra.

While the eddy--mean flow interaction of both zonal and non-zonal
perturbations is dominated by the same dynamics when $\b\ll1$, for
$\b\gg1$ the eddy--non-zonal flow interaction is dominated by
resonances which do not occur for zonal  flow perturbations. The resonant interactions lead to the possibility of arrested
decay of the vorticity flux at the rates of $\b^{-1/2}$ and $\b^{-1}$,
instead of the $\b^{-2}$ decay in the absence of resonances. The
vorticity flux attains its maximum at an intermediate value
$\b\sim\mathcal{O}(1)$ for non-zonal mean flows as well, which is
nonetheless large enough for the resonant contributions to reinforce
the contribution from the shearing mechanism. Figure \ref{fig:F_order1}
shows the contribution to the vorticity flux feedback induced by the
various wave components that are excited for two values of $\b$ ($\b=2$
and $\b=12$) in the case of zonal jets ($\varphi=0\deg$) and non-zonal
perturbations ($\varphi=15\deg$). As $\b$ increases, the resonant
contributions start playing an important role for non-zonal
perturbations as there is enhanced contribution to the vorticity flux
feedback in the vicinity of the $\Db=0$ curves, indicated by the white
dashed lines. These resonant contributions enhance the vorticity fluxes
relative to the fluxes obtained for zonal jets and render the non-zonal
structures more unstable compared to zonal jets when $\b \gtrsim
3.5$~\citep{Bakas-Ioannou-2014-jfm}.

\section{Effect of anisotropic forcing on S3T instability\label{sec:anisotr}}

In this section we investigate the effect of the anisotropy of the
excitation on the S3T instability. The maximum vorticity flux feedback
$\fr$ for three cases of anisotropy ($\mu=\pm1$ and $\mu=1/4$) and for
isotropic forcing ($\mu=0$) is shown in
Fig.~\ref{fig:frmax}\hyperref[fig:frmax]{a}. For $\b\gg1$, the main
contribution to $\fr$ for zonal jet perturbations, comes from forced
waves with nearly meridional constant phase lines (angles near
$\thet=\gamma=0\deg$, cf.~Fig.~\ref{fig:F_z_b200}). Therefore, the vorticity
flux feedback $\fr$, attains larger (smaller) values for a stochastic
forcing that injects more (less) power in waves with angles near
$\gamma=0\deg$, that is for positive (negative) anisotropicity factor
$\mu$ (cf.~Fig.~\ref{fig:spec_forc}). The maximum value of $\fr$ over
all wavenumbers $n$ depends in this case linearly on $\mu$ (cf.~\hyperref[sec:asymptotics]{Appendix~B}),
\be
\f_{r,\textrm{max}} = (2+\mu) \b^{-2} +
\Ocal(\b^{-4})\ .
\ee
For non-zonal perturbations, the main contribution comes from forced
waves satisfying the resonant condition $\om_\kv+\om_\nv=\om_{\kv+\nv}$
and $\fr$ depends only on the sum of the resonant contributions. The
sign of $\Ncal_j$ that determines whether the resonant contribution is
positive or negative (cf.~\eqref{eq:frR}), depends only on the sign of
$\sin\thet_j+n/2$ and not on the anisotropicity factor $\mu$
(cf.~\eqref{eq:defNcal}). The anisotropicity affects only
the magnitude of $\Ncal_j$. For any $0<\varphi<90\deg$ it is found that the
resonances giving positive contribution occur at angles $\thet_j$ for which
$|\gamma_j|=|\thet_j-\varphi|<45\deg$. A stochastic excitation, which injects more
power near $\gamma=0\deg$ ($\mu>0$) will give larger positive resonant
contributions and therefore $\fr$ increases with $\mu$. However, the effect on the
maximum vorticity feedback is weak, as the spectral selectivity of the
resonances renders the characteristics of the most unstable non-zonal
structure independent of the spectrum of the forcing. That is, the
$(n,\varphi)$ that correspond to the maximum $\fr$ asymptotes to
$n\approx 0.5$, $\varphi\approx 10\deg$ (marked with star in
Fig.~\ref{fig:R_Fcal}\hyperref[fig:R_Fcal]{a}) as $\b\to\infty$, a
result that is very weakly dependent on $\mu$
(cf.~Figs.~\ref{fig:frmax}\hyperref[fig:frmax]{b,c}).

For $\b\ll1$, the characteristics of the S3T instability are dependent on the anisotropy of the
stochastic forcing. The vorticity flux feedback is at leading order proportional to $\mu$:
\be
\fr =\frac1{8} \mu\,n^2 \left(1-n^2\right) \cos (2 \varphi ) + \Ocal(\b^2)\ .
\ee
This shows that there can be upgradient vorticity fluxes leading to S3T instability for $\b = 0$ as long
as $\mu \cos ( 2 \varphi) >0$. For $\mu>0$, the maximum $\fr= \mu/32$ is achieved by zonal jets ($\varphi=0\deg$),
while for $\mu<0$ any non-zonal perturbation
with $\varphi>45\deg$ can grow, with the maximum $\fr=|\mu| / 32$ achieved for $\varphi=90\deg$ when the non-zonal perturbations assume the
form of jets in the $y$ direction (meridional jets)
(cf. Fig.~\ref{fig:frmax}\hyperref[fig:frmax]{c}).


It is worth noting that \citet{Srinivasan-Young-2014} also find that
that the eddy momentum fluxes  are proportional to $\mu$ when  a constant shear  flow is  stochastically forced
with power spectrum (\ref{eq:Qhat}). This result
is intriguing as the two studies address two different physical regimes. This work treats the limit
appropriate for emerging structures in which the shear time is far larger than the
dissipation time-scale  with the fluxes  determined by the instantaneous response of the eddies on
the shear. \citet{Srinivasan-Young-2014} study the opposite limit in which
the mean flow shear is finite and the shear time is much shorter than the dissipation time-scale
with the fluxes determined by the integrated influence of the shear on the eddies over
their whole life cycle, which may include complex effects such as reflection and absorption
at critical levels.

In summary:
\begin{enumerate}
\item[a.] The S3T instability of the homogeneous state is a monotonically increasing
function of $\mu$ for all $\b$.
\item[b.] The forced waves that contribute most to the instability are structures with small $\gamma$, i.e., waves with phase
lines nearly aligned with the $y$ axis, as in Fig.~\ref{fig:spec_forc}\hyperref[fig:frmax]{a}.
\item[c.] The anisotropy of the excitation affects prominently the S3T stability of the homogeneous state
only for $\b\lesssim 3.5$.
\end{enumerate}

\section{Discussion}\label{sec:discussion}

In this work we addressed the  dynamics underlying the onset of the S3T instability leading to the formation of large scale structure but not the nonlinear development and equilibration of the instability. The emergent structure may be susceptible to either hydrodynamic or structural secondary instabilities as it reaches finite amplitude (cf.~\citet{Farrell-Ioannou-2003-structural,Farrell-Ioannou-2007-structure,Parker-Krommes-2014-generation} for zonal jets and~\citet{Bakas-Ioannou-2014-jfm} for non-zonal flows). For example, the most unstable jet structure for marginally unstable parameters  is at the scale of the forcing $L_f$ (for small $\b$ maximum instability occurs at a scale slightly larger than the forcing scale $L_f$, for isotropic forcing,
 and  close to $\sqrt{2}\,L_f$, for anisotropic forcing\footnote{The $n$  that produces  maximum instability has a non-uniform limit as $\mu \rightarrow 0$ , $\b \rightarrow 0$ because  isotropic forcing  ($\mu=0$) is singular in that the S3T homogeneous equilibrium is always stable for $\b=0$.}   while for larger $\b$  maximum instability occurs at about $2 L_f$  (cf.~Fig.~\ref{fig:frmax}\hyperref[fig:frmax]{b})).
Most  geophysical flows   are far from marginal stability and the jet scale predicted at marginal stability does not characterize the scale of the actual jets.
S3T theory predicts that the emergent jets through a series of mergers usually equilibrate at a much larger scale   \citep{Farrell-Ioannou-2007-structure,Parker-Krommes-2014-generation}.
These predictions of the S3T theory have been shown to be accurately reflected in sample nonlinear
simulations \citep{Srinivasan-Young-2012,Constantinou-etal-2014}.


Although in this work we examined the statistical dynamical instability  of a homogeneous state of turbulence
in the presence of forcing and dissipation, the results bear a relation to the deterministic barotropic hydrodynamic instability
of non-zonal flows on a  $\beta$-plane in the absence of forcing and dissipation.
\citet{Parker-Krommes-2014-book} have recently shown that in the inviscid limit the  modulational instability of a Rossby wave $\psi_\pv = A\,
\cos ( \pv\cdot\xv-\om_\pv t )$~\citep{Lorenz-1972,Gill-1974,Connaughton-etal-2010}  and
the S3T instability of a homogeneous turbulent state  with equilibrium vorticity power spectrum corresponding to the
Rossby wave: $\hat{C}^e(\kv)= (2 \pi)^2 p^4 | A |^2\[\d(\kv-\pv) + \d(\kv+\pv)\]$ obey the same stability condition.
This equivalence
is formal because physically the two problems are very different. In the problem of~\citet{Lorenz-1972},
the stability of a basic state in the form of a coherent Rossby plane wave is studied, while S3T addresses
the statistical stability of an incoherent homogeneous state with  the power spectrum of
the Rossby wave. In that sense, as noted  also by \citet{Parker-Krommes-2014-book}, S3T stability analysis
embeds the modulational instability results into a more general physical framework. In \hyperref[app:MI]{Appendix~C} we extend the result
of \citet{Parker-Krommes-2014-book} and show
the formal equivalence between the modulational instability of any solution of the barotropic equation, which
may be in general time dependent but has stationary power spectrum, with the S3T instability of the
homogeneous state with the same power spectrum. Such a nonlinear solution of the inviscid barotropic vorticity
equations is for example a superposition of any number of Rossby waves:
\be
\psi = \sum_{\substack{j=1\\ |\pv_j|=p}}^N \hspace{-0.5em} A_{j}\,
\underbrace{ \cos(\pv_j\cdot\xv - \om_{\pv_j}t) }_{\psi_{\pv_j}}\ ,
\label{eq:C1}
\ee
all with the same total wavenumber, $|\pv_j|=p$, that forms a  non-dispersive structure moving westwards
(cf.~\hyperref[app:MI]{Appendix~C}). If we assume a zonal jet perturbation superimposed on this nonlinear solution, then
the results in this work show further that the dynamics underlying the instability of this
structure can be interpreted in the limit of $\beta\ll1$ as shearing of the finite amplitude
solution by the weak shear of the jet perturbation.

\section{Conclusions\label{sec:concl}}

The mechanism for formation of coherent structures in a barotropic
beta-plane under a spatially homogeneous and temporally delta correlated
stochastic forcing was examined in this work within the framework of
Stochastic Structural Stability Theory (S3T). Within this framework, a
second order closure for the dynamics of the flow statistics is obtained
by ignoring or parameterizing the eddy--eddy nonlinearity. The resulting
deterministic system for the joint evolution of the coherent flow and of
the second order turbulent eddy covariance admits statistical
equilibria.

For a spatially homogeneous forcing covariance, a homogeneous state with
no mean coherent structures is such an equilibrium solution of the S3T
dynamical system. When a critical energy input rate of the forcing is
exceeded, this homogeneous equilibrium is unstable and propagating
non-zonal coherent structures and/or stationary zonal jets emerge in
agreement with direct numerical simulations. In order to identify the
processes that lead to the formation of coherent structures, the
vorticity fluxes induced by a plane wave mean flow, which is the
eigenfunction of the linearized S3T system around the homogeneous
equilibrium, were calculated close to the bifurcation point and closed
form asymptotic expressions for these fluxes were obtained. Upgradient
fluxes in this limit are consistent with S3T instability and coherent
structure formation.

The induced fluxes were calculated in a rotated frame of reference, in
which the plane wave mean flow corresponds to a zonal jet evolving in a
beta-plane with a non-meridional planetary vorticity gradient. This was done
because in this rotated frame of reference the intuition gained by
previous studies for the eddy--mean flow dynamics underlying zonal jet
formation can be utilized to clarify the dynamics underlying non-zonal
wave formation, or formation of zonal jets when the effect of topography
is equivalent to a non-meridional planetary vorticity gradient.

In the limit of a weak planetary vorticity gradient ($\b\ll1$), the
eddy--mean flow dynamics are similar for both zonal jets and non-zonal
structures. The stochastically forced eddies that propagate with low small
group velocities in this limit, are rapidly dissipated as they are
sheared over by the infinitesimal mean flow. Their effect on the mean
flow is therefore determined at leading order by the instantaneous, with
respect to the shear time scale, change in their momentum fluxes and to
second order by the instantaneous change in their group velocity. The
waves with constant phase lines that form angles $|\thet|<30\deg$ with
the meridional direction instantaneously surrender momentum to the mean
flow and lead to upgradient fluxes that reinforce the mean flow for an
anisotropic forcing. For an isotropic forcing this leading order effect
produces no net fluxes when integrated over all forced waves and the
instability is controlled by the second order effect that the
instantaneous change of the waves's group velocity has on the momentum
fluxes. In this case, the group velocity of waves that form angles
$|\thet|>30\deg$ with the meridional direction is instantaneously
increased (decreased) for waves propagating northward (southward) due to 
refraction. The difference in momentum fluxes resulting from this change in 
group velocity is positive in the region of their excitation leading to 
upgradient fluxes. As a result, the anisotropy of the
forcing has a significant effect on the induced fluxes and the S3T
instability in this limit. In any case, the effect of the eddies on the
mean flow due to shearing is larger for zonal jets compared to non-zonal
perturbations and consequently zonal jets are more unstable in this
limit.

In the limit of strong planetary vorticity gradient $\b\gg1$, the
eddy--mean flow dynamics producing upgradient vorticity fluxes
are different for zonal and non-zonal perturbations, but in both cases
the fluxes decrease with $\b$. Zonal jets continue
to induce upgradient vorticity fluxes through wave shearing which decrease as
$\mathcal{O}(\b^{-2})\ll1$. The reason is that in this limit
the waves that can propagate in the meridional direction
are influenced by the integrated shear over the sinusoidal flow, which is very small. However, the
non-zonal mean flow perturbations can sustain fluxes that decrease only as
$\mathcal{O}(\b^{-1/2})$. The reason for these larger fluxes is that resonant and
near resonant interactions dominate the dynamics in this
limit (cf.~section 3.26 in \cite{Pedlosky-1992}). Resonance occurs between the emerging
structure, which close to the stability boundary satisfies the Rossby wave
dispersion, and the stochastically forced waves satisfying the Rossby
wave frequency resonant condition. The resonant interactions
which occur for non-zonal structures may
produce upgradient or downgradient net fluxes and it was found that upgradient fluxes
cannot be induced by non-zonal flows with wavenumbers in a region of
wavenumber space in the shape of a dumbbell. Maximum upgradient fluxes occur
for both zonal and non-zonal flows for $\b\sim\Ocal(1)$. In this regime, shearing of the forced waves by the
infinitesimal non-zonal flows is reinforced by fluxes from the resonant interactions, enhancing
the vorticity fluxes and rendering the non-zonal structures more unstable compared to zonal jets
when $\b \gtrsim 3.5$. In contrast to the limit $\b\ll1$, these results
were found to be insensitive to the anisotropy of the forcing.

Finally, the relation of the S3T instability and
modulational instability of finite-amplitude Rossby waves was discussed.
\citet{Parker-Krommes-2014-book} showed that the growth rates obtained
when three Rossby waves interact with the primary finite-amplitude
Rossby wave, match exactly in the inviscid limit the growth rates
obtained by the S3T stability analysis for the homogeneous equilibrium
with the vorticity covariance produced by the primary Rossby wave. It
was shown in this work that this agreement can be found for more general cases
(for example when the covariance is produced by any linear combination
of Rossby waves with the same total wavenumber). Such an agreement occurs
because retaining only the interaction between four waves in
modulational instability is equivalent to neglecting the eddy--eddy
nonlinearity in S3T. The equivalence of the dynamics underlying
modulational and S3T instability in this case, shows that S3T stability analysis
generalizes modulational instability analysis in a
stochastically forced and dissipated flow. However, contrary to modulational instability, the underlying S3T dynamics
can capture both the emergence of large-scale structure and its equilibration. In addition, the
dynamics underlying modulational instability can be interpreted under
the alternative eddy--mean flow view adopted in this work.

\acknowledgments
Nikolaos Bakas is supported by the AXA Research Fund and Navid Constantinou acknowledges the support of
the Alexander S. Onassis Public Benefit Foundation. Nikolaos Bakas would also like to thank
Izumi Saito for fruitful discussions on the results of this work.

\appendix[A] 

\appendixtitle{Eddy vorticity flux response to a mean flow perturbation}
\label{app:solve_dC}


In this appendix we study the eigenvalue problem~\eqref{eq:dS3TR} which determines the S3T stability of jet perturbations to the homogeneous turbulent
equilibrium~\eqref{eq:ZeCe} in the rotated frame of reference. The eigenfunction corresponding to eigenvalue $\sigma$ has the spatial
structure:
\begin{subequations}\begin{align}
\dU &= e^{\i n y'}\ ,\label{eq:dU}\\
\dC &= \tilde{C}^{(\textrm{h})}_\nv(\xv_a-\xv_b) \, e^{\i n (y'_a+y'_b)/2}\ .\label{eq:dS}\end{align}\label{eq:dU_dS}
\end{subequations}
 The power spectrum of the homogeneous part of the covariance eigenfunction, $\tilde C^{(\textrm{h})}_\nv(\xv_a-\xv_b)$,
is determined from~\eqref{eq:dS3TR_dC} to be:
\begin{align}
\hat{C}^{(\textrm{h})}_\nv (\kv)&= \frac{\i \,\varepsilon\, k'_x}{2} \[\bit k^2_{-}\(k^2_{+}-n^2\)\hat{Q}_+ - k^2_{+}\(k^2_{-}- n^2\)\hat{Q}_- \] \times\nonumber\\
&\quad\ \times\[ \bit (\sigma+2)k^2_{+}k^2_{-} + 2\i n\b\cos{\varphi} \,k'_x k'_y -\right.\nonumber\\
&\quad\qquad\left.\bit-  \i n \b \sin{\varphi} \({k'}_x^2-{k'}_y^2+n^2/4\)\]^{-1}~,\label{eq:S_n_la}
\end{align}
with $\kv_\pm=\kv\pm\nv/2$, $\nv=(0,n)$, $k=|\kv|$, $k_\pm=|\kv_\pm|$,
$\hat{Q}'_\pm = \hat{Q}'(\kv_\pm)$ and $\hat{Q}'$
the Fourier transform of the forcing covariance~\eqref{eq:QhatR}.
The vorticity flux, $\dvz$, induced by this eigenfunction is
\begin{align}
\dvz &=\[ \frac1{2}(\Del^{-1}_a\partial_{x'_a}\!+\!\Del_b^{-1}\partial_{x'_b}) \dC\]_{\xv_a=\xv_b}
\hspace{-.8em} \nonumber\\
&=\i n\[ e^{\i n (y'_a+y'_b)/2}\]_{\xv_a=\xv_b} \times\nonumber\\
&\qquad\times\iintinf \frac{\df^2\kv}{(2\pi)^2} \[ \,\frac{k'_x k'_y}{k^2_+k^2_-} \hat{C}^{(\textrm{h})}_\nv(\kv)\,
e^{\i\kv\cdot (\xv_a-\xv_b)}\]_{\xv_a=\xv_b} \nonumber\\
& =\dU  \iintinf
\frac{\df^2\kv}{(2\pi)^2} \,\frac{\i n\,k'_x  k'_y}{k^2_+k^2_-} \,\hat{C}^{(\textrm{h})}_\nv (\kv)  \equiv \varepsilon\,\f(\s)\,\dU~,\label{eq:dvz_Lp_Lm}
\end{align}
which is proportional to $\dU$. By using the symmetry $C^e(\xv_a,\xv_b)=C^e(\xv_b,\xv_a)$, which implies that
$\hat{C}^e(\kv)=\hat{C}^e(-\kv)$, and by changing the integration variable in~\eqref{eq:dvz_Lp_Lm} to $\kv-\nv/2$, we
obtain the following expression for the feedback factor, $f$:
\begin{align}
\f(\s) &= \iintinf \frac{\df k'_x\,\df k'_y}{(2\pi)^2} \[ \bit 2n {k'_x}^2 (k'_y+n/2)\(k^2 -n^2\)\]\times \nonumber\\
&\qquad\times \[\bit (\sigma+2)k^2_{s}k^2 + 2\i n\b\cos{\varphi}\, k'_x (k'_y+n/2)\right.\nonumber\\
&\qquad\left.\bit-  \i n \b \sin{\varphi} \( {k'_x}^2- {k'_y}^2 - n k'_y\) \]^{-1}\times  \frac{{\hat{Q}'}(k'_x,k'_y)}{2}~,\label{eq:s_def}
\end{align}
with $\kv_s=\kv+\nv$ and $k_s=|\kv_s|$.

Introducing~\eqref{eq:dvz_Lp_Lm} into~\eqref{eq:dS3TR_dZ} we obtain the stability equation~\eqref{eq:S3T_nphi}
that determines the eigenvalue $\s$, which can be shown to be exactly the stability equation
obtained by~\citet{Bakas-Ioannou-2014-jfm}. The stability equation can be written in terms of the real and
imaginary part of $\s$ as:\begin{subequations}
\begin{align}
\s_r&=-1+\e\,\real[f(\s)] \ ,\label{eq:s3t_slanted_real}\\
\s_i&=\b\sin\varphi/n+\e\,\imag[f(\s)]\ .\label{eq:s3t_slanted_imag}
\end{align}\label{eq:s3t_slanted}\end{subequations}
The real part of the vorticity flux feedback $\real[f(\s)]$ contributes to the growth rate of the mean flow and
the imaginary part $\imag[f(\s)]$ determines the departure of the phase speed of the mean flow from the Rossby
wave frequency $- \beta \sin \varphi / n$. For $\b\gg1$ the first term in
\eqref{eq:s3t_slanted_imag} is $\mathcal{O}(\b)$ while for marginally unstable eigenfunctions $\imag(f)$ is at
most of $\mathcal{O}(1/\b)$. As it will be shown, the critical $\e$ increases as $\b^{1/2}$ or as $\b$ and therefore
the frequency of the marginally unstable waves is approximately equal to the Rossby phase frequency.

We focus on the real part of the feedback gain, $\real(f)$, near marginal stability ($\s_r\approx 0$). Setting $\s=-\i \om_\nv=\b\sin\varphi/n$ in (\ref{eq:s3t_slanted_real}) for the marginally unstable structures and for the ring forcing in the rotated frame, $\hat{Q}'(k'_x,k'_y) = 4\pi\,\d(k-1)\,\Gcal(\thet-\varphi)$,
 $\fr\equiv\real\[f(-\i\om_\nv)\]$ takes the
form:
\begin{align}
\fr &= \real\left(\spa\right.\int\limits_{0}^{2\pi} { \frac{\Ncal}{ \Dcal_0 +
\i\b\,\Db }}\df \vartheta \left.\spa \right)=
\int\limits_{0}^{2\pi} \frac{\Ncal\,\Dcal_0}{ \Dcal_0^2 + \b^2\,\Db^2 }\df \vartheta\ ,\label{eq:dvz}
\end{align}
 with
\begin{subequations}
\begin{align}
\Dcal_0(\thet,n)&=2(1 +n^2+2n\sin{\thet})~,\\
\Db(\thet,n)&=(1 +n^2+2n\sin{\thet})\sin{\varphi}/n+\nonumber\\
&\qquad+n^2\cos{(\thet-\varphi)} + n \sin{(2\thet-\varphi)}~,\\
\Ncal(\thet,n)&= \frac1{\pi}\, n\(1-n^2\)\,\cos^2{\thet} (\st+n/2) \, \mathcal{G}(\thet-\varphi) ~.\label{eq:defNcal}\end{align}\label{eq:defD0DbN}\end{subequations}
 Positive values of $\fr$  indicate that the  vorticity flux induced by the stochastic forcing at marginal stability  on the mean flow with wavenumber $n$ and non-zonality
parameter
$\varphi$ is upgradient, and the  marginal energy input rate is  $\e_c = 1/\fr$.

Note that   $F(\thet,n)$, which is defined in~\eqref{eq:defF} as the r.h.s. of \eqref{eq:dvz},
is unchanged when the angle $\varphi$ is shifted by $180\deg$ ($\varphi\to180\deg+\varphi$) or when there is a simultaneous shift of $\varphi\to180\deg-\varphi$ and $\thet\to180\deg-\thet$. As a result, it suffices to only consider cases with $0\le\varphi\le 90\deg$.

As \citet{Parker-Krommes-2014-book} noted, the stability equation~\eqref{eq:S3T_nphi} can be written in coordinate independent form as:
\begin{align}
&\s +1 +\i \om_{\nv} = \nonumber\\
&\ =\varepsilon  \iintinf \frac{\df^2 \kv}{(2\pi)^2}\;\frac{|\mathbf{k}\times{\mathbf{n}}|^2\, (k^2_s-k^2) (k^2-n^2) }{k^4 k^2_s n^2  \[ (\s+2) -\i\(\om_{\kv}- \om_{\kv+\nv}\)\bit \] }\frac{\hat{Q}(\kv)}{2} ~,\label{eq:sigma_nz}
\end{align}
where $\om_{\kv}$ is the Rossby frequency of a wave with wavenumber $\kv$ (defined in~\eqref{eq:def_omRossby}).
As a result, in coordinate free form
\be
\b\,\Db(\thet,n)= k^2_s (-\om_{\nv}-\om_{\kv}+\om_{\kv+\nv})\ ,\label{eq:D2zero}
\ee
and the roots of $\Db$ on the $(\thet,n)$ plane satisfy the resonant condition:
\be
\label{eq:res}
\om_{\nv}+\om_{\kv}=\om_{\kv+\nv}\ .
\ee

 \appendix[B] 

\appendixtitle{Asymptotic expressions for the induced vorticity flux feedback}
 \label{sec:asymptotics}

In this Appendix we calculate in closed form asymptotic expressions for the vorticity flux feedback
induced by a mean flow perturbation in the form of a zonal jet in the rotated frame of reference with wavenumber $n$,
 for $\b\ll1$ and $\b\gg1$.

\subsection{Case $\b\ll 1$ \label{sec:asympt_smallb}}

When $\b\ll1$ and for $n$ satisfying $\b /n\ll1$, we expand
$\Fcal(\thet,n) =F(\thet,n)+F(180\deg+\thet,n)$ in~\eqref{eq:fr} in powers of $\b$. Since $\Fcal$ is a function of
$\b^2$ we have the expansion:
\be
\Fcal=\Fcal_{0} + \b^2\, \Fcal_2 + \Ocal(\b^4)\ ,
\ee
with $\Fcal_2 = \frac1{2}\left . \bit \partial^2_{\b\b} \Fcal \right|_{\b=0}$. The leading order term is:
\begin{align}
\Fcal_0 & = \frac{n^2\(1-n^2\)}{\pi}\,\Gcal (\thet-\varphi)\frac{1+n^2-4\sin^2\thet }{ (1 + n^2)^2-4n^2\sin^2{\thet}}\cos^2{\thet} \,,\label{eq:F0}
\end{align}
due to the property $\Gcal(180\deg+\thet)=\Gcal(\thet)$. Positive values
of $\Fcal_0$ indicate that the stochastically forced waves with phase
lines inclined at angle $\thet$ with respect to the $y'$  direction,
induce upgradient vorticity fluxes to a mean flow with wavenumber $n$
when $\b=0$. Given that $n<1$ and $\Gcal>0$, $\Fcal_0$ is positive for
any forcing distribution, only in the sector shown in
Fig.~\ref{fig:smallb_F}\hyperref[fig:smallb_F]{a} in which
$4\sin^2\thet<1+n^2$. Specifically, in the absence of $\b$ all waves
with $|\thet|\le 30\deg$ reinforce mean flows with $n<1$. Note that
the condition $4\sin^2\thet<1+n^2$ is also the necessary condition for
modulational instability of a Rossby wave with wavevector components
$(\cos \thet, \sin \thet)$ to any mean flow (zonal or non-zonal) of
total wavenumber $n$ for $\b\ll 1$~\citep{Gill-1974}.

\vspace{1em}

The total vorticity flux feedback $\fr$ for
$\Gcal(\thet-\varphi) = 1+\mu\cos{[2(\thet-\varphi)]}$ is at leading order:
\be
\fr =\frac{\mu}{8} n^2 \left(1-n^2\right) \cos (2 \varphi ) + \Ocal(\b^2)\ ,\label{eq:dvz_b0}
\ee
which is proportional to the anisotropy factor, $\mu$. The maximum feedback factor is in this case
\be
\f_{r,\max} = \frac{|\mu|}{32} \ ,\label{eq:frmax_smallb_mu}
\ee
and is achieved for mean flows with $n=1/\sqrt{2}$. This maximum is achieved for
zonal jets ($\varphi=0\deg$) if $\mu>0$ and for meridional jets ($\varphi=90\deg$) if $\mu<0$.
This implies that for $\b\ll 1$ the first structures to become unstable are zonal jets if
$\mu>0$ and meridional jets if $\mu<0$, as shown in Fig.~\ref{fig:frmax}\hyperref[fig:frmax]{c}.

\begin{figure*}[t]
\centerline{\includegraphics[width=5in]{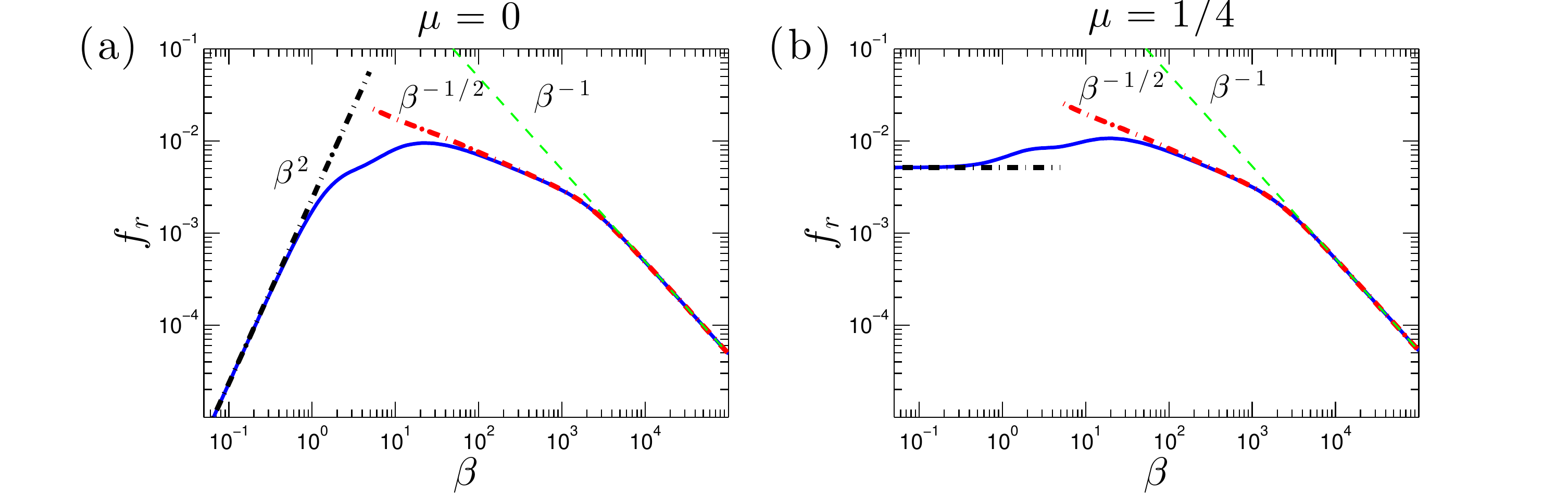}}
\appendcaption{B1}{Feedback factor $\fr$ for a non-zonal perturbation
with $n=0.4751$ and $\varphi=10^\circ$ (which belongs in region~A of Fig.~\ref{fig:R_Fcal}\hyperref[fig:R_Fcal]{a}) (solid lines) in the case of a forcing covariance with
(a) $\mu=0$ and (b) $\mu=1/4$. Also shown are asymptotic expressions for $\b\ll1$ (\eqref{eq:frmax_smallb_mu0} in (a) and~\eqref{eq:frmax_smallb_mu} in (b)) and the resonant contribution~ \eqref{eq:frR} for $\b\gg1$ (dash-dot). For $\b\gg1$, expression~\eqref{eq:fR_b} is also plotted (dashed). It can be seen that only~\eqref{eq:frR} can captures the $\b^{-1/2}$ decrease of $\fr$.}\label{fig:Rdvz}
\end{figure*}

For isotropic forcing ($\mu=0$), the leading order term is zero and $\fr$ depends quadratically on $\b$:
\begin{align}
\fr &= \b^2\frac{n^4}{64} \[\bit 2 + \cos (2 \varphi)\]+ \Ocal(\b^4)~ ~~{\textrm{for}~~~n<1},\label{eq:dvz_b0_isotr}
\end{align}
producing upgradient fluxes for $n<1$. Note that for the delta function ring forcing
$\int_0^{2\pi}\Fcal_2\,\df\thet$ is discontinuous at $n=1$, with positive values for $n=1^-$ and negative values
for $n=1^+$. The accuracy of these asymptotic expressions is shown in Fig.~\hyperref[fig:Rdvz]{B13}.

The maximum feedback factor, shown in Fig.~\ref{fig:frmax}\hyperref[fig:frmax]{a}, is
\begin{align}
\f_{r,\max} = & \frac{3\b^2}{64} \ ,\label{eq:frmax_smallb_mu0}
\end{align}
and is attained by zonal jets ($\varphi=0\deg$) with wavenumber
$n\to1^{-}$ as $\b\rightarrow 0$, a result that was previously derived by \citet{Srinivasan-Young-2012}. The
accuracy of~\eqref{eq:frmax_smallb_mu} and~\eqref{eq:frmax_smallb_mu0} extends to $\b\approx 0.1$, as shown
in Fig.~\ref{fig:frmax}\hyperref[fig:frmax]{a}.

\subsection{Case $\b\gg 1$\label{sec:asympt_largeb}}

When $\b\gg1$, we write~\eqref{eq:dvz} in the form:
\begin{align}
\label{eq:B7}
\f_r &=\frac{I} {\b^2}\ ,~~{\textrm{with}}~~I = \int\limits_{0}^{2\pi} F_\chi(\thet,n) \,\df\thet\ ,
\end{align}
where
\begin{align}
F_\chi(\thet,n) = \frac{\Ncal\,\Dcal_0}{\chi^2 \Dcal_0^2 + \Db^2 }\ ,\label{eq:refFchi}
\end{align}
and $\chi\equiv1/\b$. When $\Db\sim\mathcal{O}(1)$ for all angles $\thet$, then the feedback factor
is $\fr\sim\mathcal{O}(\b^{-2})$. However, if $\Db\sim\mathcal{O}(\b^{-1})$ for some angle $\thet$, then
as we will show in this Appendix, $\fr$ decays as $\mathcal{O}(\b^{-1})$ or as $\mathcal{O}(\b^{-1/2})$. This
is illustrated in Fig.~\hyperref[fig:Rdvz]{B13} showing the feedback factor $\fr$ as a function of $\b$ in cases
in which $\Db$ vanishes.

$\Db$ can have at most 4 roots, $0\deg \le \thet_j \le 360\deg$ ($j=1,2,3,4$), for any given $(n,\varphi)$. At these angles
the resonance condition (\ref{eq:res}) is satisfied. To calculate asymptotic approximations to the integral $I$,
we split the range of integration to a small range close to the roots of $\Db$ for which we have resonance, $I^{(\textrm{R})}$, and to a range away from the roots of $\Db$, $I^{\textrm{(NR)}}$:
\be
I = \sum_{j=1}^{N_r}\[ \spb \right. \underbrace{\int\limits_{\thet_{j-1}+\d\thet}^{\thet_j-\d\thet} \!\!
F_\chi(\thet,n)\,\df \vartheta}_{I^{(\textrm{NR})}_j} + \underbrace{\int
\limits_{\thet_{j}-\d\thet}^{\thet_j+\d\thet} \!\! F_\chi(\thet,n)\,\df \vartheta}_{I^{(\textrm{R})}_j}
\left.\spb \]\ ,\label{eq:sumI}
\ee
where $N_r$ is the total number of the roots of $\Db$ and $\thet_0\equiv\thet_{N_r}$. Asymptotic approximations to the integral over the two ranges are then found separately using a proper rescaling for the regions close to the roots of $\Db$ (cf.~\citet{Hinch-1991}).

When the distance between two consecutive roots
is $|\thet_j-\thet_{j-1}|>\sqrt{\chi}$, as in the examples shown in
Figs.~\ref{fig:R_Fcal}\hyperref[fig:R_Fcal]{c,e}, then the dominant
contribution to the integral comes from the $\mathcal{O}(\chi)$ regions close to
the roots $\thet_j$, since $F_\chi(\thet,n)$ close to $\thet_j$ is approximately a Lorentzian
of half-width $\mathcal{O}(\chi)$.
Therefore, choosing the range $\d \thet$ close to the roots to be
$\sqrt{\chi}\ll\delta\thet\ll1$, Taylor expanding $F_\chi(\thet,n)$
close to $\thet_j$ and rescaling $\thet=\thet_j+\chi u$ we obtain:
\be
I^{(\textrm{R})}_j=\frac1{\chi}\int\limits_{-\delta\thet/\chi}^{\delta\thet/\chi}
\frac{\Ncal_j\,\Dcal_{0,j}\,\df u}{\Dcal_{0,j}^2+\Dbj'^{\,2}\, u^2}+\mathcal{O}(\chi^{-3})\ ,
\ee
where $\Db'\equiv\partial_\thet\Db$ and the subscript $j$ denotes the value at
$\thet_j$. In the limit $\d \thet/\chi\rightarrow\infty$ we obtain:
\be
I^{(\textrm{R})}_j = \frac1{\chi}\frac{ \pi\,\Ncal_j}{|\Dbj'|}\ ,\label{eq:fR_b}
\ee
and as a result, the resonant contribution produces the asymptotic approximation:
\be
\fr^{(\textrm{R})} = \frac{1}{\b}\sum_{j=1}^{N_r}\frac{ \pi\,\Ncal_j}{|\Dbj'|}\ .\label{eq:fR_b}
\ee

\begin{figure}
\centerline{\includegraphics[width=16pc]{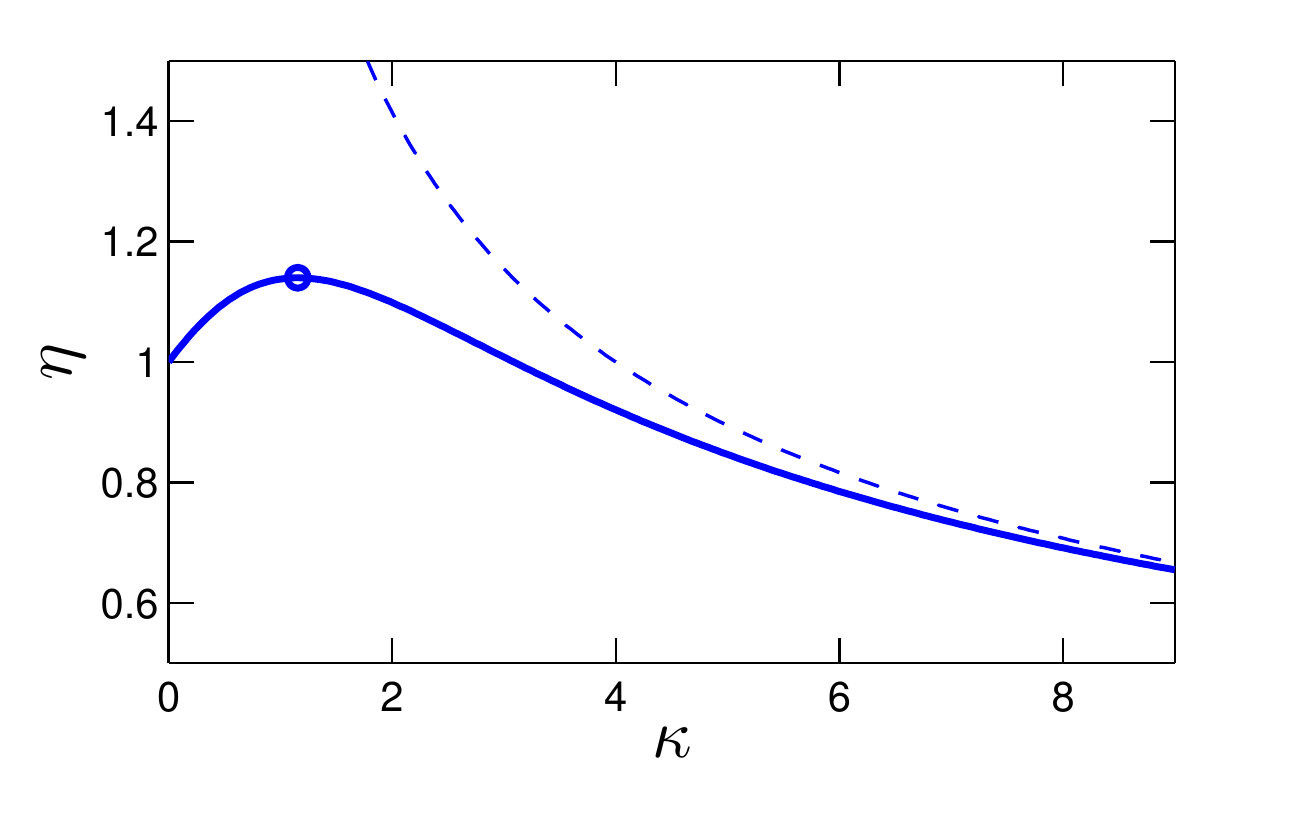}}
\appendcaption{B14}{ The factor $\eta = 2\, (\kappa^2+4)^{-3/4} \, \csc{\[\frac1{2}\arctan{(2/\kappa)}\]}$ as a function of $\kappa$ that is a measure of the distance between two consecutive resonant angles. The maximum value of $\eta$ marked with an open circle
(and consequently of the feedback gain that is proportional to $\eta$) is $\eta =3^{3/4}/2\approx 1.14$ and it is achieved at $\kappa=2/\sqrt{3}\approx1.16$. Also shown is the asymptote $\eta=2/\sqrt{\kappa}$ that $\eta$ follows for $\kappa\gg1$ (dashed). This suggests that the resonant contribution is maximum when the two roots are very close to each other ($\kappa\approx 1$) but not on top of each other ($\kappa\ll1$).}\label{fig:xifactor}


\end{figure}

However, special attention should be given to the case in which two consecutive roots are
close to each other. When $|\thet_j-\thet_{j-1}|\sim\mathcal{O}(\sqrt{\chi})$
then $\Dbj' \sim \mathcal{O}(\sqrt{\chi})$ and $\fr^{(\textrm{R})}$ scales as
$1/\sqrt{\b}$ instead of $1/\b$ for $\b\gg 1$. Indeed,
when $F_\chi$ is double peaked, as in Fig.~\ref{fig:R_Fcal}\hyperref[fig:R_Fcal]{d}, the
dominant contribution comes from the whole range between the two resonant angles which are a distance
$\mathcal{O}(\sqrt{\chi})$ apart. The proper scaling for the angles close to $\thet_j$ is therefore
$\thet=\thet_j+\sqrt{\chi}u$. Taylor expanding the denominator under this scaling we obtain:
\begin{align}
\label{eq:B13}
\chi^2\Dcal_0^2+\Db^2&=\chi^2\Dcal^2_{0,j}+\chi\Dbj'^{\,2}\, u^2+
\chi^{3/2}\Dbj'\Dbj'' u^3  +\nonumber\\
&\quad+ \chi^2\left(\frac1{4}\Dbj''^{\,2}+\frac1{3}\Dbj'\Dbj'''\right)u^4+\Ocal{(\chi^{5/2})}~,
\end{align}
where $\Dcal_{2}''\equiv \partial_{\thet\thet}^2\Db$ and $\Dcal_{2}'''\equiv
\partial_{\thet\thet\thet}^3\Db$. When $\Dbj' \sim \mathcal{O}(\sqrt{\chi})$
all the terms in (\ref{eq:B13}) are $\Ocal ( \chi^2)$ and
writing $\Dbj'=\sqrt{\x}\,d(n,\thet_j)\equiv\sqrt{\x}\,d_j$, where $d$ is of $\mathcal{O}(1)$,
the leading order resonant contribution is:
\begin{align}
&I^{(\textrm{R})}_j =\nonumber\\
&~=\chi^{-3/2} \hspace{-1em}\int\limits_{-\d\thet/\sqrt{\x}}^{\d\thet/\sqrt{\x}} \frac{\Ncal_j\,\Dcal_{0,j}\; \df u}{ \Dcal_{0,j}^2 +  d_{j}^{\,2}\, u^2 + d_{j}\la_j u^3  + \frac1{4} \la_j^{2} u^4 } + \mathcal{O}(\x^{-1}),
\end{align}
where $\lambda_j\equiv \Dbj''$. In the limit $\d\thet/\sqrt{\x}\to\infty$ the integral can be evaluated
from the residues from two of the four poles of the integrand. The poles are at $u = - d_{j}/\la_j \pm |z_j|^{1/2}\sgn{(\la_j) }\, e^{\pm\i w_j/2}$, where $|z_j| = \Dcal_{0,j} |\la_j |^{-1} (\kappa_j^2+4)^{1/2}$, $w_j=\arctan(2/\kappa_j)$ and
$\kappa_j\equiv d_{j}^2 \Dcal_{0,j}^{-1} |\la_j|^{-1}$ is an increasing function of the
distance between the two roots of $\Db$. Therefore:
\begin{align}
I^{(\textrm{R})}_j & =\chi^{-3/2} \frac{ \pi \,\Ncal_j\,\eta_j}{\Dcal_{0,j}^{1/2}|\la_j|^{1/2}}+
\mathcal{O}(\chi^{-1})\ ,\label{eq:IRj}
\end{align}
and
\be
\fr^{\textrm{(R)}} = \frac{1}{\b^2}\sum_{j=1}^{N_r}\frac1{2}\,I^{(\textrm{R})}_j= \frac{1}{\sqrt{\b}}\sum_{j=1}^{N_r}\frac{\pi \, \Ncal_j\,\eta_j}{ 2\,\Dcal_{0,j}^{1/2}|\la_j|^{1/2}}\ ,\label{eq:dvz_approx_sqrtb}
\ee
which is exactly~\eqref{eq:frR}. The factor $1/2$ in~\eqref{eq:dvz_approx_sqrtb}
arises because the range of integration includes both resonant angles and~\eqref{eq:IRj} must be divided by 2, in order to avoid double counting.
The resonant response is proportional to
\be
\eta = 2\, (\kappa ^2+4)^{-3/4} \csc{\[\frac1{2}\arctan{(2/\kappa)}\]}\ ,\label{eq:def_eta}
\ee
which is always positive, because $\kappa >0$ as $\Dcal_{0} >0$. The factor $\eta$ is shown
as a function of $\kappa$ (which is a rough measure of the distance between the roots) in
Fig.~\hyperref[fig:xifactor]{B14}. We observe that the maximum value is attained at $\kappa =2/\sqrt{3}\approx1.16$,
that is when the roots are at a distance $\mathcal{O}(\chi^{1/2})$ apart. Note also that by taking the limit
of the resonant angles being away from each other, that is by taking
the limit $\kappa\gg1$, $\eta\sim 2/ \sqrt{\kappa}$ and~\eqref{eq:dvz_approx_sqrtb} reduces to~\eqref{eq:fR_b}.
Consequently, \eqref{eq:dvz_approx_sqrtb} is a valid asymptotic expression regardless of the distance
between the roots $\thet_j$. The accuracy of~\eqref{eq:fR_b} and~\eqref{eq:dvz_approx_sqrtb} in comparison with the
numerically obtained integral is shown in Fig.~\hyperref[fig:Rdvz]{B13}.

\begin{figure*}[t]
\centerline{\includegraphics[width=.9\textwidth,trim = 40mm 0mm 40mm 0mm, clip]{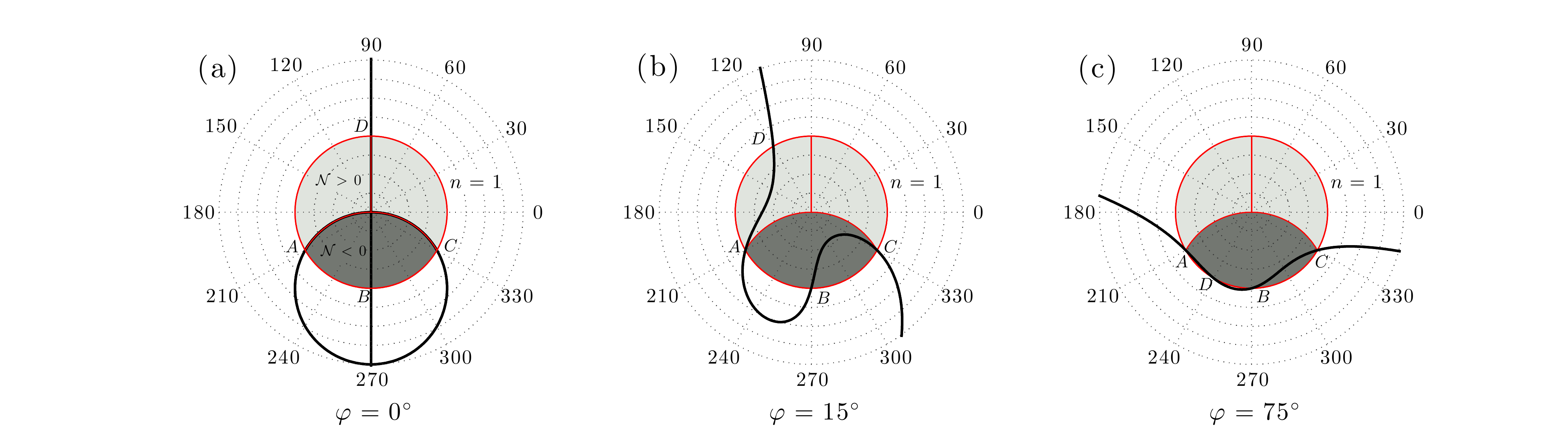}}
\appendcaption{B15}{Locus of the roots of $\Db(\thet,n)$ in a $(\thet,n)$ polar plot for (a)
zonal jet perturbations ($\varphi=0\deg$), (b) non-zonal perturbations with $\varphi=15\deg$ and (c) non-zonal
perturbations with $\varphi=75\deg$. Shaded areas indicate the region $n\le 1$. Light (dark) shade corresponds
to $(\thet,n)$ satisfying $\sin\thet>-n/2$ ($\sin\thet<-n/2$) for which we have $\Ncal>0$ ($\Ncal<0$), i.e., positive
(negative) resonant contributions. Points of intersection of the $\Db=0$ curve with the unit circle are marked with
$A$, $B$, $C$, $D$. The radial grid interval is $\Delta n=0.25$. Note that the curve $\Db=0$ does not enter the
$\Ncal>0$ area for $60\deg\le\varphi\le120\deg$, that is there are no positive contributions when
$\varphi\ge60\deg$.}\label{fig:Db}
\end{figure*}

The sign of the resonant contribution depends only on the sign of $\Ncal$. From~\eqref{eq:defNcal}
we see that $\Ncal>0$ when $\sin\thet > -n/2 $ for $n<1$; this region is highlighted with light shading in Fig.~\hyperref[fig:Db]{B15}.
It should be noted that for the important case of zonal jet perturbations ($\varphi=0\deg$) the resonant contribution is exactly zero
because $\Ncal_j=0$, as shown in Fig.~\hyperref[fig:Db]{B15a}. The asymptotic behavior of the feedback factor for this case
is found from the non-resonant part of the integral. Expanding in this case the integrand for $\chi\ll1$, we obtain
to leading order:
\be
\f_{r} \approx \f_{r}^{(\textrm{NR})}=(1-n^2)(2+\mu) \b^{-2} + \Ocal(\b^{-4})\ ,\label{eq:dvzR_largeb}
\ee
with the maximum feedback gain
\be
\f_{r,\textrm{max}} = (2+\mu) \b^{-2} + \Ocal(\b^{-4})\ ,\label{eq:sr_largeb_max}
\ee
occurring for $n\to0$. (For the special case of isotropic forcing, $\mu=0$, this reduces to the result found by \citet{Srinivasan-Young-2012}.)

Consider now non-zonal perturbations $(\varphi\ne0\deg)$. There is a large region in the $(n,\varphi)$ plane (region~D in Fig.~\ref{fig:R_Fcal}\hyperref[fig:R_Fcal]{a}) in which $\Db$ has no roots and $\fr = \Ocal(\b^{-2})$.
For larger values of $n$ (region~B in Fig.~\ref{fig:R_Fcal}\hyperref[fig:R_Fcal]{a}), and for any given $\varphi$, $\Db=0$ for exactly two $\thet_j$ that satisfy the inequality $\sin\thet_j<-n/2$. Consequently, $\Ncal_j<0$ and
the resonant contribution from these roots is negative. For even larger values of $n$ (regions A and C in
Fig.~\ref{fig:R_Fcal}\hyperref[fig:R_Fcal]{a}), $\Db$ has exactly 4 roots. Only two of the roots
in region A produce positive resonant contributions. Note also that region A extends to
$\varphi < 60\deg$ and $\varphi>120\deg$.\footnote{It can be shown that fluxes from the resonant
contributions for $n<1$ are necessarily downgradient (negative)
for $60\deg\le\varphi\le120\deg$.
Proof: A positive contribution is produced when the $\Db=0$ curve enters into the $\Ncal>0$, highlighted with light grey
in Fig.~\hyperref[fig:Db]{B15}. There are 4 roots of $\Db$ on the unit circle $n=1$ (on which also $\Ncal=0$), at angles:
$\thet=210\deg$, $270\deg$, $330\deg$ and $\thet=90\deg+2\varphi$ (marked with $A$, $B$, $C$ and $D$ respectively). The $\Db=0$ curve can cross the curve $AOC$,
 which separates positive from negative $\Ncal$, only at points $A$ and $C$, since $\Db=0$ only at these points on $AOC$.
Therefore, the $\Db=0$ curve can
enter the $\Ncal>0$ region i) through $D$, if it lies outside the arc $ABC$, and/or ii) through $A$, $C$. However, for
$60\deg\le \varphi \le 120\deg$ point $D$ lies within the arc $ABC$ and moreover, the gradient $\nablav \Db$ at points $A$ and $C$ is oriented in such way that does not allow the $\Db=0$ curve to enter $\Ncal>0$, as $\partial_n\Db<0$ and $\partial_\thet \Db\le 0$ ($\partial_\thet \Db\ge 0$) at point $A$ (point $C$).}

The maximum response, which is $\Ocal(\b^{-1/2})$, arises in region A close to
the curve separating regions~A and C where $\kappa \approx 1.16$. While the roots of $\Db$
are independent of $\b$, the location and the size of the region of maximum response
depends on $\b$ through the dependence of $\kappa$ on $\b$. However,
as $\b$ increases this dependence is weak and as $\b \rightarrow \infty$ the maximum
response occurs in a narrow region near $n \approx 0.5$ and $\varphi \approx 10\deg$, marked
with a star in Fig.~\ref{fig:R_Fcal}\hyperref[fig:R_Fcal]{a}. The width of this region decreases with $\b$, making it exceedingly
hard to locate for large $\b$, and the asymptotic approach of $(n,\varphi)$ to $(0.5,10\deg)$ is shown
in Fig.~\ref{fig:frmax}\hyperref[fig:frmax]{b,c}.

\appendix[C]
\appendixtitle{Formal equivalence between S3T instability of a homogeneous equilibrium with  modulational instability of a corresponding basic flow}
\label{app:MI}

 In this Appendix we  demonstrate the formal equivalence between the modulational instability  (MI) of any solution of the barotropic equation, which may be in general time dependent but has stationary power spectrum, with the S3T instability of the homogeneous state with the same power spectrum.

Consider a  solution $\psi_G(\xv,t)$, with vorticity $\z_G = \Delta \psi_G$, of the inviscid and unforced nonlinear barotropic equation~\eqref{eq:nl} with time-independent power spectrum. Because $J(\psi_G, \z_G)=0$,
$\z_G$ satisfies the equation
\begin{align}
\partial_t\,\z_G& =\Lcal^{\textrm{(h)}} \z_G\,\label{eq:psiG}\ ,
\end{align}
with $\Lcal^{\textrm{(h)}} =\zhat\cdot\( \bv\times\nablav\)\Del^{-1}$. Linear perturbations $\delta \zeta$ to this solution evolve according to the equation:
\begin{align}
\partial_t\,\d\z& =\Lcal\,\d\z\ ,\label{eq:dpsi_lorenz}
\end{align}
where
\begin{align}
\Lcal &= \underbrace{-\uv_G \cdot\nablav + (\Del\uv_G)\cdot\nablav\Del^{-1}}_{\Lcal_G'}+\underbrace{\zhat\cdot\( \bv\times\nablav\)\Del^{-1}}_{\Lcal^{\textrm{(h)}}}\nonumber\\
&=\Lcal_G' + \Lcal^{\textrm{(h)}}\ ,\label{eq:C4}
\end{align}
is the time-dependent linear operator about $\z_G$ that has been decomposed into a spatially homogeneous
operator, $\Lcal^{\textrm{(h)}}$, that governs the evolution of $\z_G$ and the inhomogeneous operator
$\Lcal_G'$ that depends on $\z_G$.  The hydrodynamic instability of  $\z_G$ is ascertained when
the largest Lyapunov exponent of (\ref{eq:dpsi_lorenz}) is positive.

We proceed with the study of the modulational instability by  decomposing the perturbation into a mean $\d Z=\<\d \zeta\>$ and deviations from the mean
$\d \zeta '=\d \z-\d Z$, where $\<\;\bullet\;\>$ is an averaging operation. The averaging operation
in modulational instability  is projection to the eigenstructure with
wavenumber $\nv$, which is orthogonal to $\z_G$, because only orthogonal   eigenstructures to $\zeta_G$ could become unstable.
With this averaging  operator
$\< \zeta_G \> =0$, and therefore $\z_G=\z'_G$, whereas the perturbations
has a non-zero mean, $\d Z$, and a deviation and is expressed as  $\d \z = \d Z +\d \z'$.
For example, if $\psi_G$ is  a sum of Rossby waves as in~\eqref{eq:C1} the
perturbation field from Bloch's theorem comprises of Fourier components with wavenumbers
$\nv,~\nv\pm \pv_j,~\nv \pm2\pv_j,~\nv\pm3\pv_j,\dots$ for all  the $\pv_j$. In this case $\d Z$ is a plane wave
with wavenumber $\nv$ and $\d \z'$ comprises of the remaining Fourier components.
With these definitions ~\eqref{eq:dpsi_lorenz} is equivalently written as:
\begin{align}
\partial_t\(\d Z+\d\z'\)& =\Lcal'_G \d Z + \Lcal^{\textrm{(h)}}\d\z' + \Lcal'_G\, \d\z' + \Lcal^{\textrm{(h)}} \d Z\ ,\label{eq:dZ+dz}
\end{align}
where $\Lcal'_G $ is primed in order to stress that the operator linearly depends on the
deviation quantity $\z'_G$. Equation~\eqref{eq:dZ+dz} is then separated to form an equivalent system of
equations for the evolution of the mean perturbation, $\d Z$, and the deviation perturbation, $\d\z'$:
\begin{subequations}
\label{eq:system}
\begin{align}
\partial_t\,\d Z& = \Lcal^{\textrm{(h)}} \d Z + \< \Lcal'_G\d\z' \> \ ,\label{eq:MIdZ}\\
\partial_t\,\d\z'& = \Lcal^{\textrm{(h)}}\d\z' +\Lcal'_G\,\d Z +\Lcal'_G\,\d\z'- \< \Lcal'_G\,\d\z' \>\ .\label{eq:MIdz}
\end{align}\end{subequations}
The stability equation (\ref{eq:dpsi_lorenz}) and the stability equations (\ref{eq:system}) for $\d Z$ and
$\d \z'$ are equivalent. In modulational instability studies the term $\Lcal'_G\,\d\z'- \< \Lcal'_G\,\d\z' \>$ in
(\ref{eq:MIdz}) is neglected and the
stability of the following simpler system is studied:
\begin{subequations}
\label{eq:systemMI}
\begin{align}
\partial_t\,\d Z& = \Lcal^{\textrm{(h)}} \d Z + \< \Lcal'_G\d\z' \>\ , \label{eq:MI1}\\
\partial_t\,\d\z'& = \Lcal^{\textrm{(h)}}\d\z' +\Lcal'_G\,\d Z\ .\label{eq:MI2}
\end{align}\label{eq:MIdz_trunc}\end{subequations}
For example, if $\psi_G$ is in the form of~\eqref{eq:C1} the  neglected term comprises waves with wavevectors $\nv\pm2\pv_j,~\nv\pm3\pv_j,\dots$
and the truncated system~\eqref{eq:MIdz_trunc} allows only  interaction between the primary finite amplitude waves $\pv_j$, the perturbation $\nv$
and the waves $\nv\pm\pv_j$. If  $\z_G$  is a single wave $\pv$ (as in MI studies),
\eqref{eq:systemMI} is referred to as the 4 mode truncation or `4MT' system.

However, instead of studying the MI stability of $\d Z$ and $\d \z'$ using the approximate~\eqref{eq:systemMI} equations, we can equivalently study the stability of $\d Z$ and $\d C(\xv_a,\xv_b,t) = \<\bit\right. \z'_{G}(\xv_a,t)\,\d\z'(\xv_b,t)+\z'_{G}(\xv_b,t)\,\d\z'(\xv_a,t)\bit \left.\bit\> \equiv \<\bit\right.\z'_{G,a}\,\d\z'_b+\z'_{G,b}\,\d\z'_a \left.\bit\>$.
With these definitions we obtain from (\ref{eq:psiG}) and (\ref{eq:MI2}) the evolution equation for $\d C$:
\begin{align}
\partial_t \d C &= \< (\partial_t\z'_{G,a})\,\d\z'_b + (\partial_t\z'_{G,b})\,\d\z'_a \right.+\nonumber\\
&\qquad\left.+ \z'_{G,a}\,(\partial_t\d\z'_b)+ \z'_{G,b}\,(\partial_t\d\z'_a) \bit \>\nonumber\\
 &=  \(\Lcal^{\textrm{(h)}}_a +\Lcal^{\textrm{(h)}}_b\)\d C +\nonumber\\
&\qquad+ \< \z'_{G,a}\, \Lcal'_{G,b}\,\d Z_b +  \z'_{G,b}\, \Lcal'_{G,a}\,\d Z_a \bit \>~.\label{eq:c8}
  \end{align}
We note from the definition of $\Lcal'_G$ (cf.~\eqref{eq:C4}) that:
\begin{align}
\Lcal'_{G}\,\d Z 
&= -\(\zhat\times{\bm\nabla\psi'_G}\) \cdot\nablav \d Z+ \(\zhat\times{\bm\nabla\z'_G}\)\cdot\nablav\d \Psi\nonumber\\
&= \(\bit\zhat\times{\bm\nabla\d Z}\) \cdot\nablav \psi'_G - \(\bit\zhat\times{\bm\nabla\d\Psi}\)\cdot\nablav \z'_G\nonumber\\
 & = (\Del\,\d\Uv)\cdot(\nablav \psi'_G) - (\d\Uv)\cdot(\nablav\z'_G) = \d\Acal \;\z'_G\ ,
\end{align}
where $\d\Uv=\zhat\times\nablav\d\Psi$ is the velocity field associated with $\d Z$ and $\d\Acal = -\d\Uv\!\cdot\!\nablav + (\Del\;\d\Uv)\!\cdot\!\nablav\Del^{-1}$ is the operator that also
appears in~\eqref{eq:s3t_pert_dC}. As a result
(\ref{eq:c8}) becomes:
\begin{equation}
\partial_t \d C = \(\Lcal^{\textrm{(h)}}_a +\Lcal^{\textrm{(h)}}_b\)\d C + \(\d\Acal_a +\d\Acal_b\bit\) C^G\ ,
\end{equation}
where $C^G= \< \z'_{G,a} \z'_{G,b} \>$. Returning now to (\ref{eq:MI1}) we note that
$ \< \Lcal'_G\d\z' \> = \Rcal(\d C) $, where $\Rcal(\d C)$ is defined in~\eqref{eq:defR}, as:
\begin{align}
&\Rcal( \d C ) =\nonumber\\
&\ \ =-\nablav\cdot\[ \frac{\zhat}{2}\times(\nablav_a\Del^{-1}_a+\nablav_b\Del^{-1}_b)\right.\nonumber\\
&\hspace{9em}\left.\vphantom{\frac{\zhat}{2}} \< \bit \z'_{G,a}\,\d\z'_b+\z'_{G,b}\,\d\z'_a \>\]_{\xv_a=\xv_b}\nonumber\\
&\ \ =-\nablav\cdot\left\{\vphantom{\frac1{2}} \zhat\times\< \bit (\nablav\psi'_{G})\d\z'  + (\nablav\d\psi') \z'_G\>\right\}\nonumber\\
&\ \ =-\nablav\cdot\< \bit \uv'_G\,\d\z'  + \d\uv'\, \z'_G\>\nonumber\\
&\ \ =\<\bit  -\uv'_G\cdot\nablav\,\d\z'  +  (\Del\uv'_G)\cdot\nablav \d\psi'\> = \< \Lcal'_G\,\d\z' \>\ .
\end{align}

Consequently, the MI of $\z'_G$ is equivalently determined from  the stability of the system:
 \begin{subequations}
\label{eq:system1}
\begin{align}
\partial_t\,\d Z& = \Lcal^{\textrm{(h)}} \d Z + \Rcal(\d C)\ ,\label{eq:M1}\\
\partial_t \d C & = \(\bit\right.\Lcal^{\textrm{(h)}}_a +\Lcal^{\textrm{(h)}}_b\left.\bit\)\d C + \(\d\Acal_a +\d\Acal_b\bit\) C^G\ ,
\end{align}\end{subequations}
 which is identical to equations~\eqref{eq:s3t_dZdC} that determine the S3T stability of the homogeneous equilibrium with zero mean flow, $\Uv^e=0$, and
equilibrium covariance $C^e=C^G$ under the ergodic assumption that ensemble averages are equal to averages under operation $\< \;\bullet\; \>$.

For example, consider the nonlinear solution
\be
\psi(\xv,t) = \int\limits_0^{2 \pi} a(\thet) \cos ( \pv \cdot \xv - \om_{\pv}t) \,\df \thet\ ,
\label{eq:C10}
\ee
with wavevectors $\pv=(\cos\thet,\sin\thet)$ on the unit circle ($p=1$). Expanding the plane waves into
cylindrical waves:
\be
e^{\i\[ ( x+\b t)\cos\thet +y\sin\thet \]} = \sum_{m=-\infty}^{+\infty} \i^m J_m(\varrho) e^{\i m (\phi-\thet)}\ ,
\ee
with $\varrho^2 = (x+\b t)^2 + y^2$, $\phi = \arctan\[y/(x+\b t)\bit\]$ and $J_m$
the $m$-th Bessel function of the first kind, this can be shown to be the non-dispersive
structure
\be
\psi (x+\b t,y) = \real\[ \sum_{m=-\infty}^{+\infty} \gamma_m\,J_m(\varrho) e^{\i m \phi} \]\ ,
\ee
propagating westward with velocity $\b$, where
$\gamma_m = \int_0^{2\pi} a(\thet)\,e^{-\i m \thet}\,\df\thet$. The results in this
Appendix show that the modulational instability of the propagating structure
\eqref{eq:C10} is equivalent to the S3T instability of the homogeneous equilibrium with
covariance $C^e$ prescribed by power spectrum
$\hat{C}^e(\kv) = (2\pi)^2 \left| a(\thet)\right|^2\,\d(k-1)$. Note that this S3T equilibrium is  also an exact  homogeneous statistical equilibrium of the
nonlinear barotropic equations without approximation.


%


\end{document}